%% file: BA_BABApaper1.tex
\RequirePackage{etoolbox}
\csdef{input@path}{%
 {sty/}
 {img/}
}%
\csgdef{bibdir}{bib/}

\documentclass[ba]{imsart}
\pubyear{0000}
\volume{00}
\issue{0}
\doi{0000}
\firstpage{1}
\lastpage{1}

\usepackage{amsthm}
\usepackage{amsmath}
\usepackage{natbib}
\usepackage[colorlinks,citecolor=blue,urlcolor=blue,filecolor=blue,backref=page]{hyperref}
\usepackage{graphicx}

\usepackage{epsfig}
\usepackage{xypic}

\usepackage{natbib}
\usepackage{acronym}
\usepackage{enumerate}
\usepackage{color}

\usepackage{soul}
\usepackage{amsmath,amsfonts,amssymb}
\usepackage[all]{xy}
\usepackage{psfrag}
\usepackage{psfrag}
\usepackage{epsfig}
\usepackage[all]{xy}
\usepackage{multirow}

\startlocaldefs

\newcommand{\e}[1]{\ensuremath{{\rm E}[#1]}}

\newcommand{\var}[1]{\ensuremath{{\rm Var}[#1]}}

\newcommand{\cov}[2]{\ensuremath{{\rm Cov}\left[#1,#2\right]}}

\newcommand{\be}{\begin{equation}}
\newcommand{\ee}{\end{equation}}
\newcommand{\ba}{\begin{eqnarray}}
\newcommand{\ea}{\end{eqnarray}}
\newcommand{\bi}{\begin{itemize}}
\newcommand{\ei}{\end{itemize}}
\newcommand{\bn}{\begin{enumerate}}
\newcommand{\en}{\end{enumerate}}
\newcommand{\bfi}{\begin{figure}}
\newcommand{\efi}{\end{figure}}

\newcommand{\SD}[1]{\text{SD}[#1]}

\endlocaldefs

\begin{document}


\begin{frontmatter}
\title{A Bayesian computer model analysis of \\Robust Bayesian analyses}

\runtitle{A Bayesian computer model analysis of Robust Bayesian analyses}

\begin{aug}
\author{\fnms{Ian} \snm{Vernon}\thanksref{addr1}\ead[label=e1]{i.r.vernon@durham.ac.uk}}
\and
\author{\fnms{John Paul} \snm{Gosling}\thanksref{addr2}\ead[label=e2]{j.p.gosling@leeds.ac.uk}}

\runauthor{}

\address[addr1]{Department of Mathematical Sciences, Durham University, Science Laboratories, Durham, DH1 3LE, UK,
 \printead{e1}
}
\address[addr2]{School of Mathematics, University of Leeds, Leeds, LS2 9JT, UK,
 \printead{e2}
 }


\end{aug}

\begin{abstract}
We harness the power of Bayesian emulation techniques, designed to aid the analysis of complex computer models, to examine the structure of complex Bayesian analyses themselves. These techniques facilitate robust Bayesian analyses and/or sensitivity analyses of complex problems, and hence allow global exploration of the impacts of choices made in both the likelihood and prior specification. 
We show how previously intractable problems in robustness studies can be overcome using emulation techniques, and how these methods 
allow other scientists to quickly extract approximations to posterior results corresponding to their own particular subjective specification.
The utility and flexibility of our method is demonstrated on a reanalysis of a real application where Bayesian methods were employed to capture beliefs about river flow. We discuss the obvious extensions and directions of future research that such an approach opens up.

\end{abstract}


\end{frontmatter}

\section{Introduction}

Bayesian methodology is now widely employed across many scientific areas \citep[for example, over 100 articles have been published in Nature with the word ``Bayesian" in the title;][]{nature16}. 
The uptake of Bayesian methods is due both to progress in Bayesian theory and to advances in computing power combined with the development of powerful numerical algorithms, such as MCMC.
However, many Bayesian analyses of real world problems are both complex and computationally time-consuming. They often involve complex hierarchical models that require large numbers of structural and distributional assumptions both in the likelihood and prior (along with other choices covering the numerical implementation). Due to the long run times and the need to tune such algorithms, it is common for little or no rigorous sensitivity analysis to be performed, therefore it is often unclear as to what extent the Bayesian posterior and the subsequent decisions it informs have been affected by these numerous assumptions. For any serious scientific analysis, a solid understanding of the inferential process and its response to changes in the underlying judgements and assumptions is absolutely vital. Any Bayesian analyses that cannot do this is of limited use and, we would assert, has questionable worth to the scientific community. 

Much work has been done to address the issues of robustness and sensitivity analysis of Bayesian analyses, with many elegant results derived \citep[see for example][]{box1,berger1,Berger:2000aa,roos1}. 
However, progress in this area has greatly slowed over the past fifteen years due in part to the intractability of analysing even fairly basic Bayesian models. In particular, although aspects of prior sensitivity were explored \citep[see e.g.][]{berger1,Moreno:2000aa,Fan:2000aa} and loss sensitivity \citep[][]{Dey_Insuabook}, perturbations to the likelihood proved far more challenging to deal with analytically~\citep{Shyamalkumar:2000aa}. 
Two broad robust Bayesian strategies can be distinguished, the first of these being the global approach, whereby whole classes of priors and/or likelihoods are considered, and their effects on the posterior analysed. While there was much early success in this direction \citep[see for example][]{berger_pvalue_evid,berger1,Moreno:2000aa}, many of these results relied upon appeals to monotonicity arguments which were of great use in lower dimensional cases, but less easy to apply in more complex, higher dimensional models. Even defining sensible prior or likelihood classes to investigate in high dimension, while avoiding vacuous results, becomes problematic \citep{insua1}. Increasing attention was also directed at a second strategy, that of the local sensitivity approach, whereby the derivatives of posterior features of interest with respect to perturbations of various forms are analysed often using differential calculus \citep[see for example][]{Gustafson:1995aa,Gustafson_Insuabook,Perez:2005aa,Zhu_loc_sen,Muller_loc_sens,Roos_Held:2011}. While far more tractable, 
the local approach has obvious weaknesses, in that its results may be strongly sensitive to the original prior and likelihood specification. For many complex Bayesian models, for which the posterior features may be highly non-linear functions of the perturbations, such local approaches will be clearly inadequate.



Despite the efforts of the robust community, it must be conceded that the huge advances in MCMC and comparable numerical methods, which allow the use of more and more complex Bayesian models, have left robust Bayesian analysis techniques far behind \citep[][]{watson2016,robert2016}. As complex Bayesian models along with MCMC algorithms are now widely used in areas of real world importance, and as our Bayesian community will be judged upon the results of these algorithms, the need for powerful, general robust methods 
applicable to a wide class of perturbations is increasingly urgent. This article suggests a framework for the solution to this problem.

%
We propose to treat a complex and computationally demanding Bayesian analysis as an expensive computer model. We utilise Bayesian emulation technology developed for complex computer models \citep[as described in][for example]{oh2006a} to explore the structure of the Bayesian analysis itself, and, specifically, its response to various changes in both the prior and likelihood specification. This allows for a more general sensitivity and robustness analysis that would be otherwise unattainable, because we do not require analytic solutions. This methodology is very flexible, provides both local and global results, is straightforward to implement in its basic form using currently available emulation software packages, and can deal with a wide class of statistical analyses.

In more detail, a typical Bayesian analysis involves many judgements and assumptions, both in relation to modelling choices that feed into the likelihood and in terms of the representation of prior beliefs. Often, pragmatism leads to assumptions being made that are based either entirely or in part on mathematical convenience.  For example, a conjugate analysis falls into this category with convenient mathematical forms chosen in both the likelihood and prior. Aside from modelling choices, expressing judgements in probabilistic form can be time consuming and difficult, so in many cases tractable representations followed by simple assessments are made that only approximately represent the beliefs of the expert. At the other extreme, so-called objective priors are used either due to their reported properties, or perhaps because any relevant subjective beliefs are thought to be too weak to alter the analysis to any noticeable degree. All of the above compromises are defensible only if it can be shown that the posterior attributes of interest are relatively insensitive to small changes in the prior and modelling specifications. Our approach is to explore the concerns regarding the specific choices and assumptions used to form the prior and modelling specifications by embedding the current Bayesian analysis within a larger structure, constructed by parameterising the major set of choices made, following the robust Bayesian paradigm. This 
larger structure is then subjected to Bayesian computer model techniques. 
While not all choices can be parameterised directly, as we will discuss, often the major sources of concern can be addressed in this way.

Our approach also addresses another major concern: that of multiple subject area experts, who each may possess different beliefs regarding the prior and likelihood structures. 
Even when a thorough Bayesian analysis, possibly using MCMC, is performed and published, its results are usually  
based on the judgements of a single expert (or small group of experts). It is therefore difficult 
for other experts in the area to know how to interpret these results: what they really require is for the MCMC to be rerun with their beliefs inserted instead. 
Therefore, at the very least, the statistician should facilitate the analysis of a class of prior or likelihood statements, approximately representing the differing views held across the relevant scientific community. Unfortunately this is not provided in the vast majority of Bayesian analyses, albeit due to understandable constraints on time and computational resources.
However, our analysis will enable experts to quickly extract approximations to their posterior results corresponding to their own specification, along with 
associated uncertainty statements. This is what many scientific fields require: complex Bayesian analyses that are simultaneously applicable to a range of scientific specifications. 

The article is organised as follows. In Section~\ref{sec_BA_CCM} we recast the problem of a robust Bayesian analysis into that of a complex computer model, describe computer model emulation methodology, and then apply it to an example Bayesian model. In Section~\ref{sec_riverflowiv} the utility and flexibility of our method is demonstrated on a reanalysis of a real application where Bayesian methods were employed to capture beliefs about river flow. We discuss the various choices one is faced with in this kind of analysis, and outline several areas of future research in Section~\ref{sec_disc_cm}, before concluding in Section~\ref{sec_conc}.

\section{Bayesian analysis as a complex computer model}\label{sec_BA_CCM}

Our set-up is similar in structure to that of a robust Bayesian analysis; however, we utilise a computer model representation and notation~\citep[see for example][]{Craig97_Pressure,kenoh2001,Higdon04_prediction,Vernon10_CS}.
Let us assume that interest lies in a vector of random quantities $\theta$, beliefs about which will be updated in the light of a vector of data $z$. 
The prior $\pi(\theta| x_p)$ and likelihood $l(z|\theta,x_l)$ are both conditioned 
on some specific list of choices and modelling assumptions represented by parameters $ x_p$ and $ x_l$ respectively, an example of which would be hyper-parameters that have been kept constant.
We wish to explore features of interest of the posterior $\pi(\theta|z, x_p, x_l)$ such as the 
mean, variance, quantiles, etc. chosen due to their relevance to the downstream application or decision process. We map the posterior to this vector of attributes 
using the functional $g(.)$ and, hence, define the overall function $f(x)$ as:
\be\label{eq_fx}
f(x) \;\;=\;\; f(x_p,x_l) \;\;=\;\; g(\pi(\theta|z, x_p, x_l))
\ee
where $x = (x_p,x_l)$ is the combined vector of inputs that parameterise the specific choices and assumptions made in the prior and likelihood specifications, and $f(x)$ is the vector of all posterior features and summaries of interest \citep[even decision end-points if we extend the Bayesian model to include utility functions in a similar approach to][]{oakley2009}, 
where the dependence on the data $z$ is now implicit. Note that it would be simple to extend equation~(\ref{eq_fx}) to include a loss function and any
corresponding associated inputs, if necessary. An example of 
$f(x)$ that we use in Section~\ref{ssec_toy_mod_setup}, where the posterior mean and standard deviation are of primary interest is:
\be
f(x)\; =\; \left(\e{\theta| z,x},\SD{\theta| z,x}\right).
\ee

For most Bayesian analyses, in order to evaluate the posterior, we require a possibly expensive sampling algorithm such as MCMC, which may 
take hours, days or even weeks for one evaluation for a particular choice of inputs $x$.
Hence, we can view the implementation of the Bayesian analysis as an expensive computer model $f(x)$, that maps a possibly high dimensional input vector $x$ to 
a vector of outputs $f$ of primary interest to the modeller. 
Note that we would be free to view the MCMC algorithm itself as a stochastic computer model, in which case we could add any algorithm inputs $x_{\mbox{\tiny MCMC}}$ such as parameters governing the adaptive regime, burn-in and so on to the input vector $x$. We could also include additional diagnostic outputs into the vector $f$ such as the MCMC acceptance rates. However, we leave such complications to future work, as here, we are primarily interested in the key features $f(x)$ of the underlying Bayesian analysis itself, which the MCMC output only approximates. The precise representation of the link between the MCMC output and $f(x)$ will be given in Section~\ref{ssec_cmemul}.


We then seek to explore the behaviour of the posterior features of interest $f(x)$ as a function of the inputs $x$ across 
a wide class of Bayesian analyses defined as
\be
\mathcal{F} \; = \; \{f(x):x\in \mathcal{X} \}
\ee
where $\mathcal{X}$ governs the extent of our robust-Bayesian analysis and allows us to explore 
simultaneous changes in the prior and likelihood specifications. Note that in general, for a high dimensional and large enough $\mathcal{X}$, we would expect both the location and shape of the 
posterior $\pi(\theta|z, x)$ to vary substantially over $\mathcal{X}$, and hence that standard techniques based around re-sampling an individual MCMC sample \citep[see for example][]{M:1992aa,Geweke:1999aa}, or importance sampling (see for example \cite{Fan:2000aa,SINHARAY:2002aa}), may not be effective.

We envisage that the need to explore a class of Bayesian analyses may arise for several reasons: for example, we may wish to perform a global robust Bayesian analysis over 
$\mathcal{X}$ due to a possibly imprecise specification or to perform a local sensitivity analysis. Alternatively, we may be dealing with a collection of experts whose opinions on the prior and likelihood differ, but which are all contained within $\mathcal{X}$. Therefore, we depart somewhat from the goal of a typical robust analysis in that we are primarily interested in the  entire behaviour of 
$f(x)$ over the set $\mathcal{X}$, and not just in the maximum and minimum values of $f(x)$ that could be attained. This is 
because we want our results to be applicable for any user that has a precise or imprecise specification 
contained within $\mathcal{X}$, and because we may also wish to understand and identify any sensitive regions where $f(x)$ rapidly changes as a function of $x$. 
%
Unlike in many computer model analyses, we therefore do not view $x$ as being uncertain:
if this was the case we would simply build an additional layer of prior structure over $x$ into our Bayesian hierarchical model. Instead,
we seek to investigate and efficiently represent, using an emulator, the behaviour of $f(x)$ for any value of $x\in \mathcal{X}$.
If an 
expert subsequently came with their own specification $x_e$, they would instantly be able to read off the likely values of the 
posterior features of interest $f(x_e)$ corresponding to their own particular beliefs. Additionally,
the results of our analysis should provide approximate answers to any local robustness, global robustness or sensitivity analysis question regarding $f(x)$, critically, with an attached 
statement of uncertainty. The emulator structure that incorporates this uncertainty can also guide future evaluations of the sampling algorithm 
designed to resolve key uncertainties of most interest to the expert(s).
As we attempt to represent a large class of inputs and outputs, our approach is more general than a perfunctory robust Bayesian analysis, and should be widely applicable.
We now go on to describe the emulation process, and how to adapt it for application to the analysis of Bayesian analyses.

\subsection{Computer model emulation}\label{ssec_cmemul}

Emulation is a powerful technique for modelling and subsequently analysing expensive computer models that may possess high dimensional input and output spaces. Emulation has been successfully applied to complex models across a wide range of scientific disciplines 
including cosmology~\citep{Vernon10_CS,Higdon09_Coyote2,
galf_stat_sci,Vernon:2016aa}, epidemiology~\citep{Yiannis_HIV_1,Yiannis_HIV_2}, oil reservoir modelling~\citep{Craig96_Pressure,Craig97_Pressure,JAC_Handbook}, climate modelling~\citep{Williamson:2013aa,johnson2}
and environmental science~\citep{ken2008,asses_mod}.

The computer model is viewed as an expensive function that maps a vector of inputs $x$ to a vector of outputs $f(x)$. For clarity, we describe here the case for univariate $f(x)$, but multivariate emulators are constructed in similar fashion \citep[see for example,][]{Rougier:2008aa,conti1}. 
The computer model can only be evaluated at a limited number of $n$ inputs $x_D=\{x^{(1)}_D,\dots,x^{(n)}_D\}$ over the $d$-dimensional input space $\mathcal{X}$ due to restrictions on computational resources (many computer models take hours, days or even weeks to run a single evaluation, similar to current MCMC run times). Beliefs about the value of the uncertain function $f(x)$ at an untried input $x$ are represented by a Gaussian process, also termed an emulator. More formally,
\begin{equation}
  f(.)|m(.),c(.,.)  \;\; \sim \;\;  GP(m(.),c(.,.)),
\end{equation}
where $m(.)$ is a function of the model inputs $x$ that captures our beliefs about the global behaviour of the model and $c(.,.)$ is a covariance function that captures beliefs about the smoothness of the function and the overall uncertainty. A common choice in the emulation literature for the covariance function is that of Gaussian form:
\begin{equation}\label{eq_cor_gauss}
c(x,x') \;\;=\;\; \sigma_{em}^2 \exp\{-||x - x' ||^2/\theta_{em}^2 \},
\end{equation}
where $\sigma_{em}^2$ and $\theta_{em}$ are emulation parameters that need to be specified. This choice of covariance structure is 
sometimes seen as too restrictive by practitioners in the field of computer experiments due to the fact that functions sampled from the Gaussian process are infinitely differentiable, but alternative structures are available. For example, the Matern correlation function, which (in 1-dimension) is given by:
\begin{equation}\label{eq_cor_matern}
c(x,x') \;\;=\;\; \sigma_{em}^2 \;\frac{2^{1-\nu}}{\Gamma (\nu)} \left( \frac{x-x'}{\theta_{em}} \right)^{\nu} K_{\nu} \left( \frac{x-x'}{\theta_{em}} \right),
\end{equation}
where $K_{\nu}$ is a modified Bessel function of the third kind and $\theta_{em}$ and $\nu$ are parameters to be specified that govern the correlation length 
and the derivatives of the computer model respectively ($\nu$ rounded up to the next integer gives the number of derivatives that exist). 
Examination of emulator diagnostics~\citep{bastos1} can guide the choice of correlation structure.   


Updating the emulator with the vector of outputs $f(x_D)$ provides a representation of our beliefs about the behaviour of the computer model over the whole of the input space parameterised by $x$, from which we can extract say the mean and variance $\e{f(x)|f(x_D)}$ and $\var{f(x)|f(x_D)}$. Depending on the specific form of the emulator, we may have the posterior:
\begin{equation}\label{eq_GPpost}
 \!\!\!\!\!\!\!\!  f(.)|f(x_D),m(.),c(.,.) \;\; \sim \;\; GP(m^*(.),c^*(.,.)),
 \end{equation}
 where the posterior mean is given by
 \begin{eqnarray}\label{eq_GPpostmean}
 m^*(x)  \;\; = \;\; m(x) + \cov{f(x)}{f(x_D)} \var{f(x_D)}^{-1} (f(x_D)-\e{f(x_D)})
 \end{eqnarray}
and the posterior covariance is given by
 \begin{eqnarray}\label{eq_GPpostcor}
 c^*(x,x')  \;\; = \;\; c(x,x') - \cov{f(x)}{f(x_D)} \var{f(x_D)}^{-1}\cov{f(x_D)}{f(x')}.
 \end{eqnarray}
Evaluation of the emulator, in terms of its mean and variance, for different values of $x$, is very efficient and is usually several orders of magnitude faster that the original computer model. Due to the substantial gain in evaluation speed, the behaviour of $f(x)$ can be investigated far more thoroughly, and sensitivity analysis, history matching, calibration and many other powerful techniques can be performed \citep[see for example,][]{oakoh2004,kenoh2001,Vernon10_CS,Vernon10_CS_rej}.

The ability to efficiently simulate joint realisations from the emulator can also be very beneficial. For example,
often interest may lie in estimating the maximum and minimum of $f(x)$ over some subset of the input space of interest $\mathcal{X}_k \subset \mathcal{X}$. We can use a large number of joint simulations to provide estimates for the maximum and minimum of $f(x)$ over $\mathcal{X}_k$, with associated uncertainty statements, as is performed for several robust Bayesian calculations given in Section~\ref{sec_ex_spec}. These estimates will likely benefit from the smoothness assumption of $f(x)$, leading to increased accuracy. 


Another useful feature of Gaussian process emulation is its representation of derivatives. If the computer model function $f(x)$ is a Gaussian process, then the partial derivatives $\partial f(x) / \partial x_i$ also form Gaussian processes, with covariance function naturally constructed by taking the partial derivatives of $c(.,.)$ \citep{oh1992}. This is useful when we are required to provide a local sensitivity analysis around a particular Bayesian analysis, or set of analyses, which requires estimates of all the $\partial f(x) / \partial x_i$ along with corresponding uncertainty statements, or for incorporating derivative information if available.

More advanced forms of the emulator are of course possible. There is much debate in the computer model literature regarding how much 
structure should be built into the mean function $m(.)$, with some preferring constant or linear terms in all inputs \citep{oh2006a}, while 
others use more complex functions like low-order polynomials~\citep{Vernon10_CS,Vernon10_CS_rej}. In the latter case, a more advanced emulator for general output $f_i(x)$ may be represented as:
\begin{equation}\label{eq_dur_em}
f_i(x) \;\; = \;\; \sum_j \beta_{ij}  g_{ij}(x_{A_i}) + u_i(x_{A_i}) + w_i(x) 
\end{equation}
where $\beta_{ij}$ are unknown constants, the $x_{A_i}$ are the active variables, a subset of the inputs that are found to be most influential for output $f_i(x)$, 
$g_{ij}(x_{A_i})$ are known deterministic functions of the active inputs, $u_i(x_{A_i})$ is a Gaussian process now defined only over the active inputs, with mean zero and correlation function given by for example equations~(\ref{eq_cor_gauss}) or (\ref{eq_cor_matern}), and $w_i(x)$ is an uncorrelated nugget term that represents the effects of the inactive variables. Note that the use of active inputs in particular, which effectively performs an individual dimensional reduction for each output, can help overcome the many problems associated with high input dimension.
See for example~\cite{Vernon10_CS,Vernon10_CS_rej} for further discussions on emulator structure. 

When constructing emulators, various diagnostics are available to check emulator performance, see for example~\cite{bastos1}. Once an emulator has been constructed, variance-based sensitivity indices \citep{saltelli1} can be calculated efficiently using the probabilistic sensitivity analysis techniques described in \cite{oakoh2004}. The sensitivity indices can be used to give an indication of which model inputs are responsible for most variation in the model outputs (given the range of plausible values for the inputs): the main-effect indices give the proportion of variance in the output explained by a input acting on its own and the total-effect indices give the proportion of variance in the output explained by a input on its own and in conjunction with other inputs.

\subsubsection{Adapting emulation for application to a Bayesian analysis}\label{sssec_appBA}

All of the above emulation methodology can, with slight modification, be applied to the outputs of an MCMC algorithm as part of a robust Bayesian analysis, as represented by $f(x)$, with $x\in \mathcal{X}$, or indeed to any statistical analysis that is expensive to perform and for which one requires a sensitivity analysis.  

We would start by designing a space filling batch of $n$ runs $x_D=\{x^{(1)}_D,\dots,x^{(n)}_D\}$ over the $d$-dimensional input space $\mathcal{X}$. The MCMC algorithm would then be run at each of the design points, and the usual convergence tests and examination of mixing plots would be performed. 
Our framework can of course incorporate information from alternative MCMC algorithms, as we discuss in section~\ref{ssec_future}, however convergence issues may favour the approach described here.
Due to the large number of burn in steps required for MCMC convergence, a suitable design would most likely favour a smaller number of design points with a large number of posterior samples drawn at each point: a classic computer model set up. 
The computer model literature tentatively recommends at least $10d$ points in the design (with $d$ being the dimension of the input space). However, we would stress that this is highly dependent on the number of active inputs, and on the complexity of the surface one is trying to emulate: often more are needed. Here, we use space filling designs \citep[see for example,][]{morris1}, with large numbers of posterior samples, and leave a more detailed treatment of such design questions to future work. 

An important difference from the standard deterministic computer model emulation setup is that, as the MCMC algorithm only returns draws from the posterior, it should be viewed as a stochastic computer model, and hence an allowance should be made for the fact that we only see, for example, sample means and sample variances and not the true posterior values. There are many approaches to the emulation of stochastic computer models 
of varying complexity \citep[see for example][]{johnson1,Yiannis_HIV_2, Ver:stoch_sysbio1}. Here, we generate large MCMC samples and treat the resulting low level of stochasticity via a simple nugget representation as follows.  
Although simple, we make the connection between the true posterior quantities and their MCMC sample counterparts explicit to facilitate a later discussion of the partial derivatives of $f(x)$, which are of use in a local sensitivity analysis. 
Representing the Bayesian posterior features of interest as $f(x)$ and the corresponding sample quantities obtained from the MCMC algorithm as $f^{(s)}(x)$, we model the link between the two for output $i$ as:
\be
f_i^{(s)}(x) \;\;=\;\; f_i(x) + \eta_i(x)
\ee
where $\eta_i(x)$ is an uncorrelated nugget term possessing zero mean and constant variance across the input space, usually estimated from the MCMC run data~\citep{Yiannis_HIV_1}. Note that as the effective sample size of the MCMC runs will be large, the variance of 
$\eta_i(x)$ will be far smaller than other uncertainties, and more detailed modelling will be in many cases unwarranted. 

\begin{center}
\begin{figure}
\begin{tabular}{cc}
\hspace{-1.5cm} \includegraphics[scale=0.46,angle=0,viewport=-1 -1 480 360,clip]{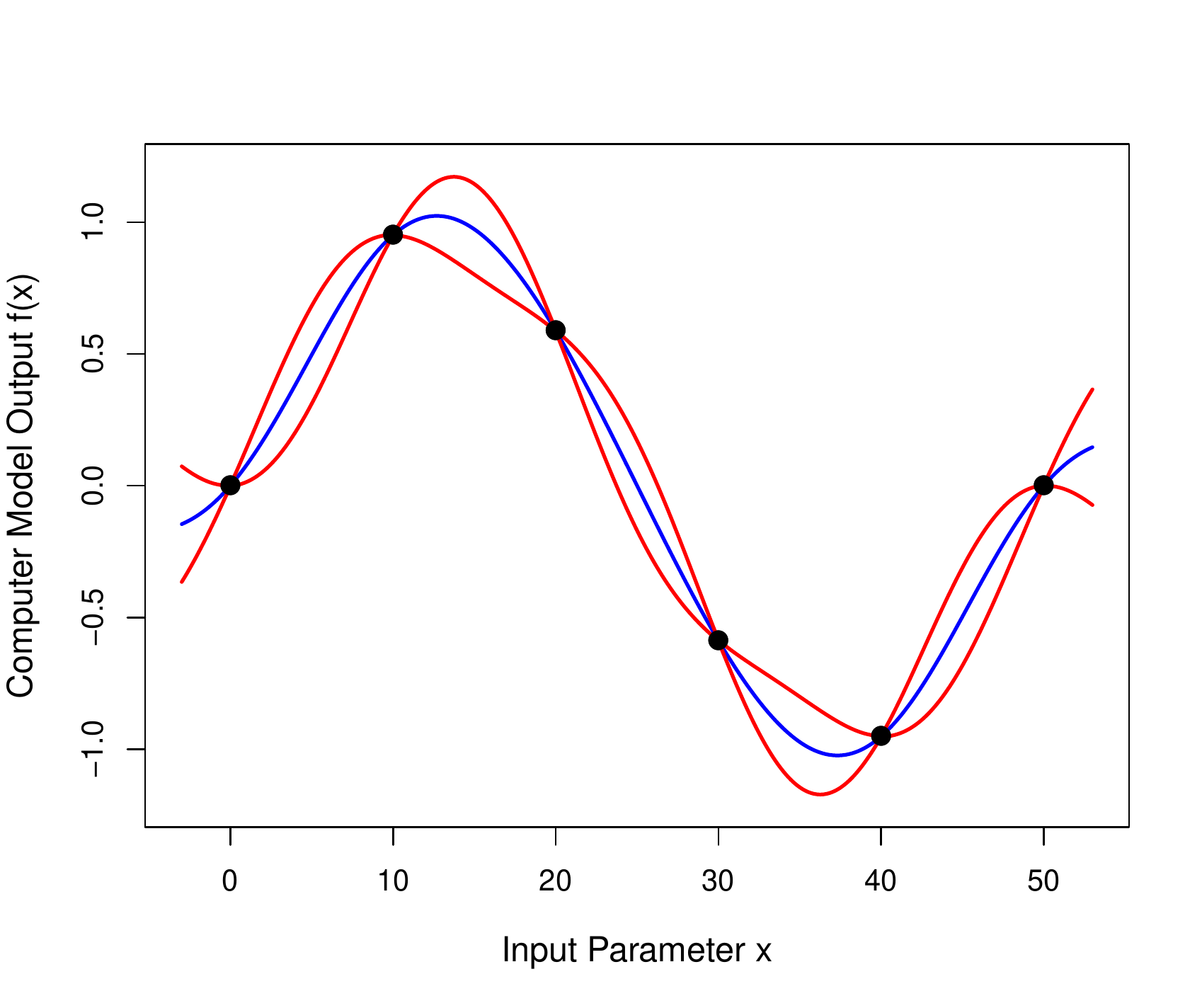} & 
\hspace{-0.6cm} \includegraphics[scale=0.46,angle=0,viewport=-1 -1 480 360,clip]{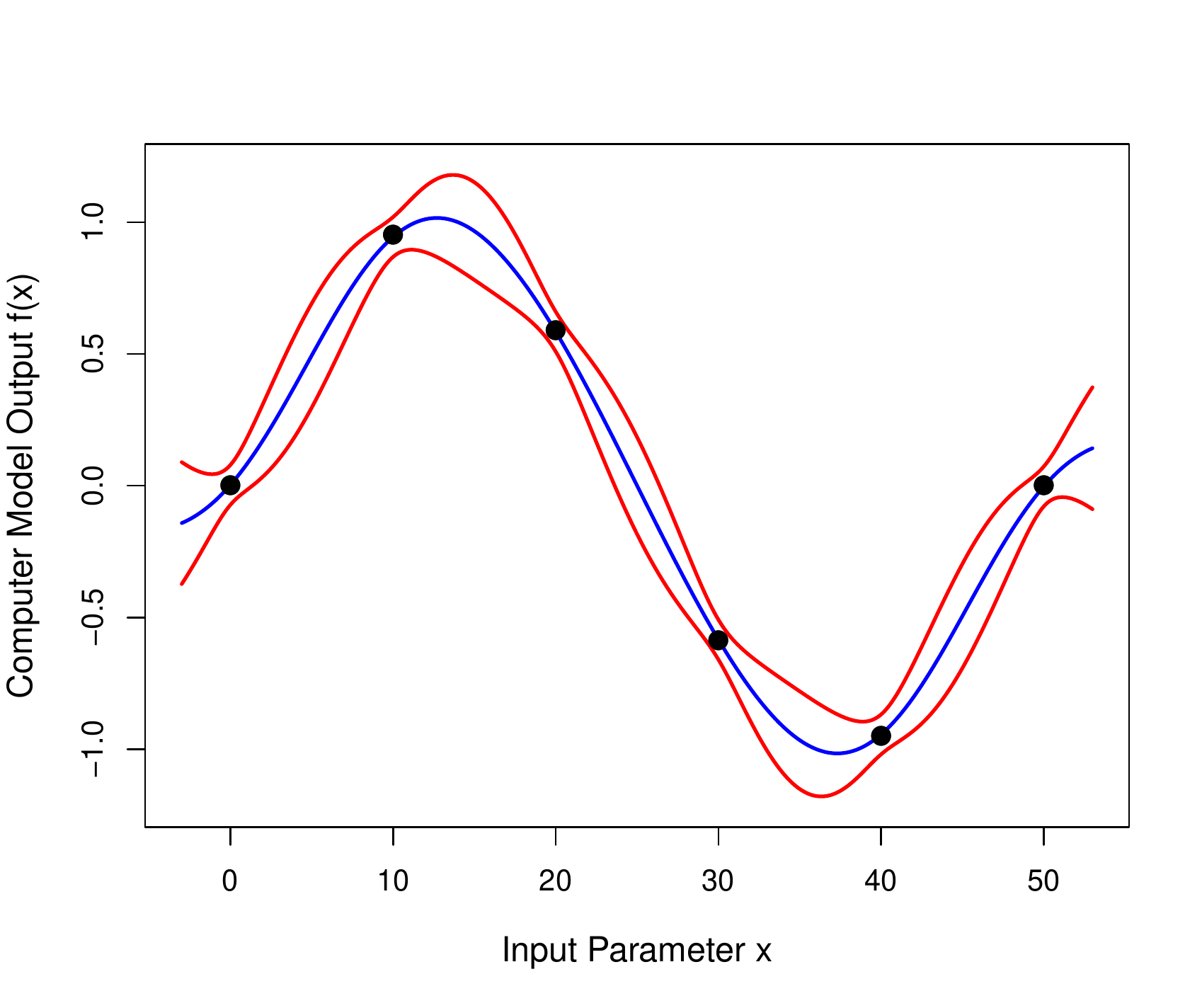} \\
\end{tabular}
\vspace{-0.5cm}
\caption{\footnotesize{Left Panel: a typical 1 dimensional emulator for a deterministic model, showing the expectation of $f(x)$ (blue line) and 95\% credible intervals (red lines) after 6 model evaluations (given by the black points). Right Panel: an emulator for a stochastic model, where the black points now represent the sample means of 6 sets of repeated model evaluations at 6 design points. Note that the emulator correctly no longer exactly interpolates the points, nor do the credible intervals shrink to zero size, in contrast to the deterministic case.}}
\label{fig_toy_emul1}
\end{figure}
\end{center}

We may believe that $f_i(x)$ is smooth and, hence, choose an appropriate correlation structure for it, given say by equation~(\ref{eq_cor_gauss}). It follows that the correlation function for the MCMC output $f_i^{(s)}(x)$ becomes
\be\label{eq_mccov}
\cov{f_i^{(s)}(x)}{f_i^{(s)}(x')} \;=\; c^{(s)}(x,x') \;=\; \sigma^2_{em} \left[ (1-\delta_{em}) \exp\{-||x - x' ||^2/\theta^2_{em} \}  + \delta_{em} \delta_{x,x'} \right]
\ee
where $ \delta_{x,x'} =1$ when $x=x'$ and 0 otherwise, $\delta_{em}$ controls the influence of the nugget variance, and $\sigma^2_{em}$ now represents the prior variance of $f_i^{(s)}(x)$. The covariance between 
$f_i(x)$ and $f_i^{(s)}(x)$ is now
\be\label{eq_mcbacov}
\cov{f_i(x)}{f_i^{(s)}(x')} \;=\; \sigma^2_{em} (1-\delta_{em}) \exp\{-||x - x' ||^2/\theta^2_{em} \} 
\ee
We can construct an emulator for $f_i(x)$ as before using the expressions for the posterior mean and correlation given by equations~(\ref{eq_GPpostmean}) and (\ref{eq_GPpostcor}), but now we replace all occurrences of $f(x_D)$ by $f^{(s)}(x_D)$ in equations~(\ref{eq_GPpostmean}) and (\ref{eq_GPpostcor}), and use 
equations~(\ref{eq_mccov}) and (\ref{eq_mcbacov}) to evaluate the altered covariance terms. Figure~\ref{fig_toy_emul1} (left panel) shows an example of a simple deterministic emulator of the function $f(x)=\sin (2\pi x/50)$, while the right panel gives the equivalent emulator of the same function observed with noise, analogous to the stochastic MCMC sample case. 

Another benefit of this construction, where we have implicitly included the smoothness of $f(x)$ (noting that $f^{(s)}(x)$ is of course not smooth), is 
that we can also construct emulators for the partial derivatives $\partial f(x) / \partial x_j$ for minimal extra computational cost. These follow the same principals, but with the correlation between the derivatives $\partial f(x) / \partial x_j$ and the MCMC outputs $f^{(s)}(x)$ now given for output $i$ by:
\be\label{eq_mcderivcov}
\cov{\frac{\partial f_i(x)}{\partial x_j}}{f_i^{(s)}(x')} \;=\; -\frac{2}{\theta^2}\sigma^2_{em} (1-\delta_{em}) (x_i-x_i') \exp\{-||x - x' ||^2/\theta^2_{em} \}   
\ee
which is obtained by partially differentiating equation~(\ref{eq_mcbacov})~\citep{oh1992}. The derivative emulators are evaluated using equations~(\ref{eq_GPpostmean}) and (\ref{eq_GPpostcor}) as before, but now with $f(x)$ replaced by $\partial f(x) / \partial x_j$.


Once the emulators have been constructed, they can be used to explore the behaviour and both the local and global sensitivity of the outputs of the Bayesian analysis $f(x)$ to the decisions made, as represented by the inputs $x$. Questions of robustness can be addressed, and various graphical methods can be employed to explore these, which we develop in Sections~\ref{ssec_toy_mod_setup} and \ref{sec_em_ba_rain}. Many other useful computer model techniques still have important analogies in this setting e.g. history matching, model discrepancy and calibration, and we discuss their uses in Section~\ref{sec_disc_cm}. We now go on in the next section to demonstrate 
our techniques on an example Bayesian model.

\newpage 
\subsection{Example Bayesian model}\label{ssec_toy_mod_setup}

We introduce a simple synthetic example of a Bayesian analysis to demonstrate the proposed methodology. Despite the simplicity of this model, it exhibits some interesting features in terms of the response of the posterior to the prior and likelihood specification, that highlight the utility of our  approach.  
We investigate a case in which we imagine there is reasonable disagreement between experts over both the prior and likelihood specifications.

Scalar data $z_i>0$ with $i=1\dots 10$ are observed, and we imagine that a Bayesian analysis has initially been performed with the following conjugate specification. The data, given in appendix A, are assumed to be independent and identically distributed with likelihood given by
\begin{eqnarray}\label{eq_toy_conj}
 z_i|\theta ~& \sim &  {\rm Exp}(\theta), \quad \quad  i=1,\dots,10 \\ \label{eq_toy_conj2}
 \Rightarrow \;\; \pi(z_i| \theta) &=& \, \theta \, e^{-\theta z_i}, \quad \quad z_i \ge 0 
\end{eqnarray}
parameterised in terms of the rate parameter $\theta$, which has corresponding prior
\be
\theta|\mu,\nu^2 ~\sim~ {\rm Ga}(\mu,\nu^2)
\ee
where Ga$(\mu,\nu^2)$ denotes a gamma distribution that has been parameterised in terms of its mean and variance, $\mu$ and $\nu^2$ respectively. Initially, the prior hyperparameters were judged to be $\mu_0=5$ and $\nu_0=1$.

Given data, this Bayesian analysis would be easy to implement given that the prior distribution is conjugate. We imagine that there is however concern amongst the experts about the data generating process, specifically with the tails of the likelihood and its behaviour close to $z_i=0$. We explore these concerns by contaminating the likelihood with a half-normal component 
$z_i| \theta \: \sim \: HN(\theta)$, where 
the impact of the contamination is controlled by a mixing parameter $\epsilon \in [0,1]$. When $\epsilon=1$ the likelihood is purely half-normal so that
\ba
z_i| \theta,\epsilon \!=\!1 ~& \sim & HN(\theta),  \quad \quad  \quad  \quad ~i=1,\dots,10 \\
\Rightarrow \;\; \pi(z_i| \theta,\epsilon \!=\!1) &=& \frac{2}{\pi} \, \theta \, e^{-\theta^2 z_i^2/ \pi}, \quad \quad z_i \ge 0  \label{eq_contam}
\ea
where we have parameterised the half-normal distribution in terms of its inverse mean 
$\theta$, such that $\e{z_i|\theta,\epsilon \!\!=\!\!1} = 1/ \theta$, in direct agreement with the definition 
of $\theta$ in the uncontaminated exponential likelihood of equations~(\ref{eq_toy_conj}) and (\ref{eq_toy_conj2}).

\begin{figure}
\begin{center}
\begin{tabular}{cc}
\hspace{-1.8cm} \includegraphics[scale=0.47,angle=0,viewport=-1 -1 415 385,clip]{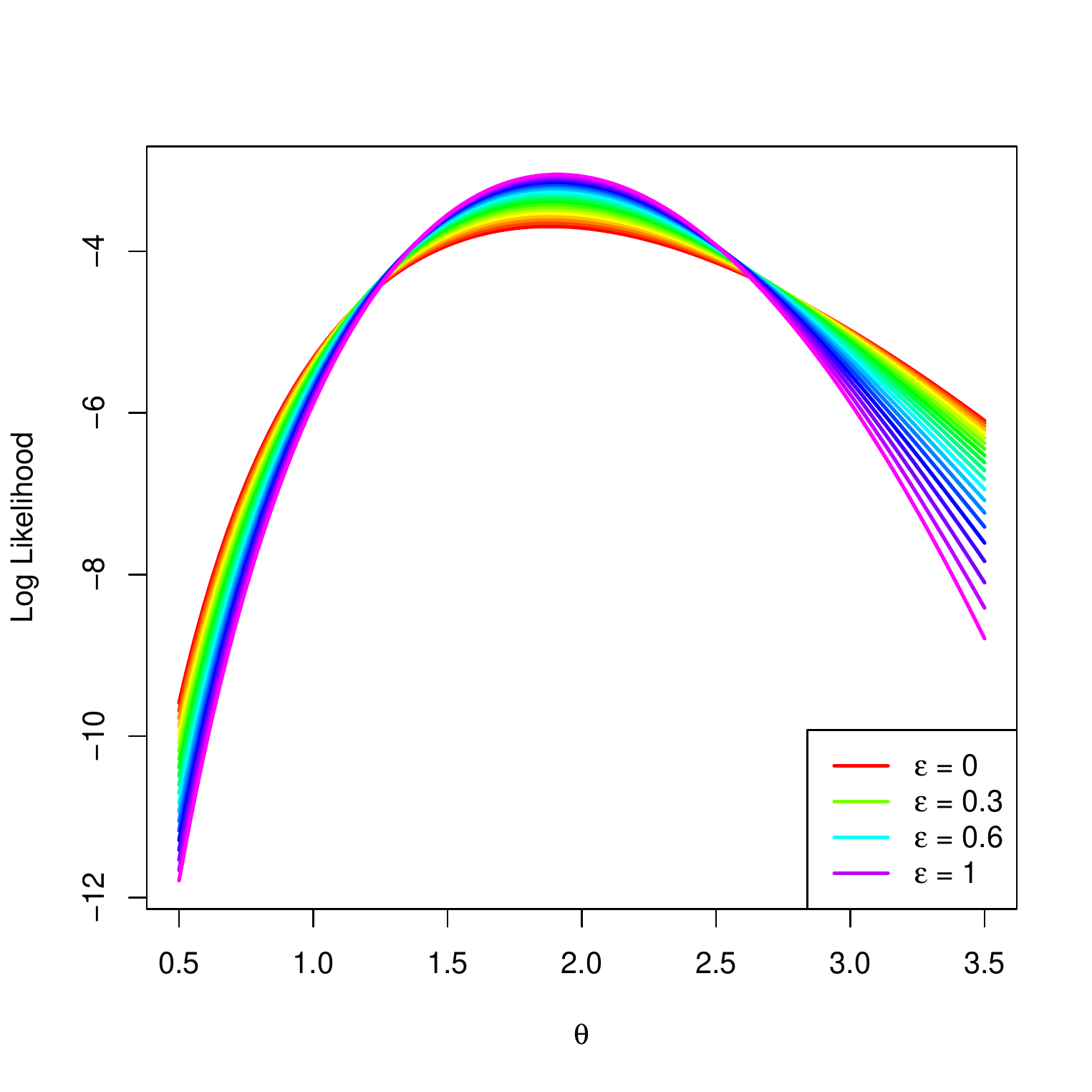} &
\includegraphics[scale=0.49,angle=0,viewport=-1 -1 505 380,clip]{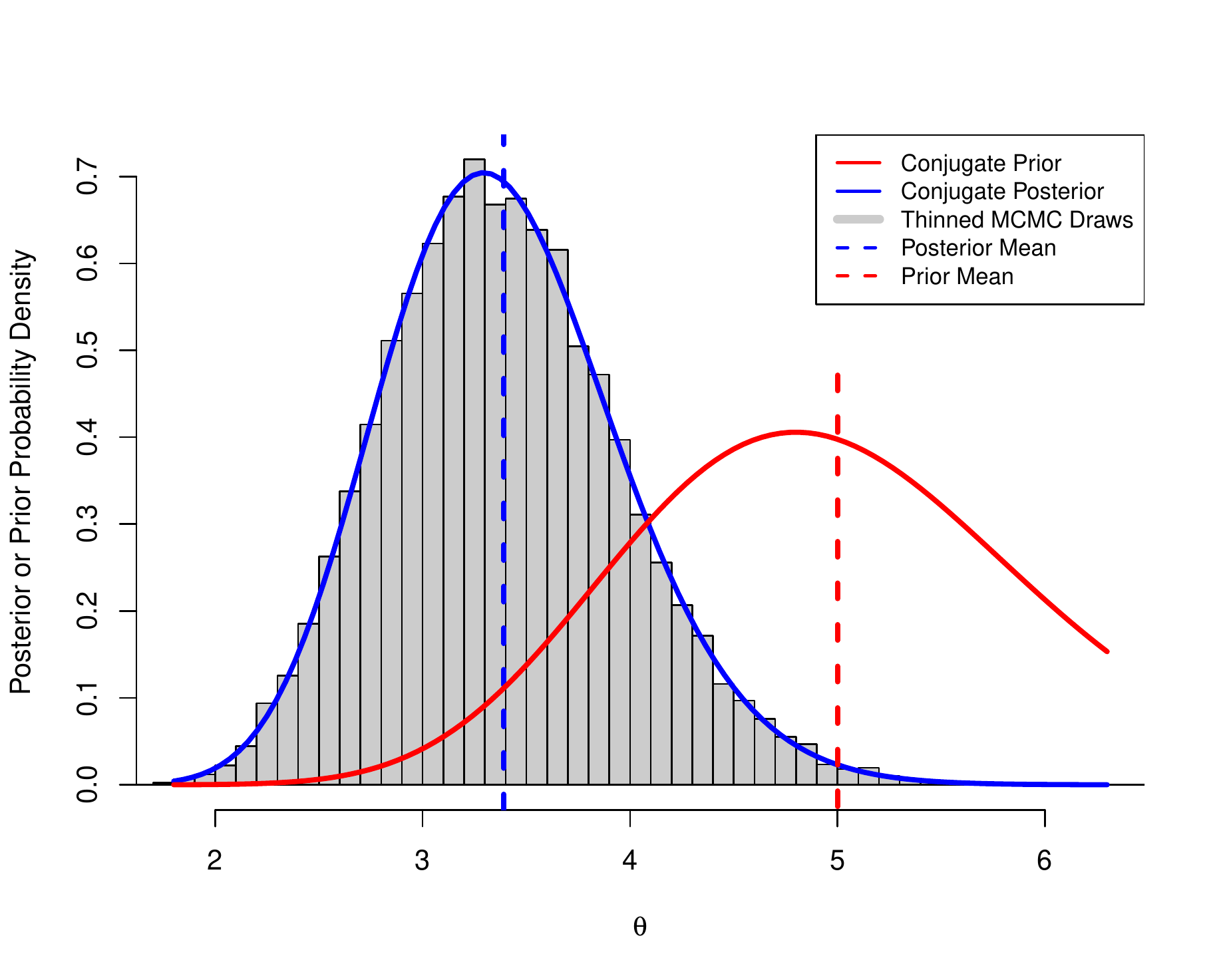}    
\end{tabular}
\vspace{-0.6cm}
\caption{\footnotesize{Left panel: the contaminated log-likelihood $\log(\pi(z|\theta,\epsilon))$, given by equation~(\ref{eq_toy_con_lik}), as a function of $\theta$, with the colours labelling differing values of the contamination parameter $\epsilon$. Right panel: draws from the MCMC algorithm in the original conjugate case (when $\nu=1$ and $\epsilon = 0$), showing the theoretical posterior density $\pi(\theta | z,\nu\!=\!1,\epsilon\!=\!0)$ in blue and the prior density $\pi(\theta | \nu\!=\!1)$ in red. The prior and posterior means are given as the vertical dashed lines in red and blue respectively, the later is the first output $f_1(x)$ to be emulated. This plot therefore represents the single point $x=(1,0)$ in the space of possible Bayesian 
analyses denoted by $\mathcal{X}$.}}
\label{fig_conjcase}
\end{center}
\end{figure}

The full contaminated likelihood for $z = (z_1,\dots,z_{10})$, conditioned on the contamination parameter $\epsilon$,
can now be written as
\begin{equation}\label{eq_toy_con_lik}
 \pi(z|\theta,\epsilon) \;\; = \;\; \prod_{i=1}^{10} \left(
		(1-\epsilon) \, \theta \,e^{-\theta z_i} + \epsilon \, \frac{2}{\pi} \, \theta \,e^{-\theta^2 z_i^2/ \pi} 
\right).
\end{equation}
where we have ensured that the property $\e{z_i|\theta,\epsilon} = 1/ \theta$ still holds for any $z_i$ and now any $\epsilon$, consistent with the original specification. Figure~\ref{fig_conjcase} (left panel) shows $\pi(z|\theta,\epsilon)$ as a function of $\theta$ for various levels of $\epsilon$.
The contamination parameter $\epsilon$ represents, and is used to investigate, the experts' disagreement over the structure of the likelihood, and returns it to the pure exponential form and hence to conjugacy as $\epsilon\rightarrow 0$.
 
The experts are still satisfied with a gamma prior and agree with the prior mean $\mu_0=5$, but not with the prior variance $\nu^2=1$ for which there is a range of alternative opinions:
\begin{equation}\label{eq_2}
 \theta|\mu_0,\nu^2 ~\sim~ {\rm Ga}(\mu_0,\nu^2), \quad \quad 0.3 < \nu < 2;
\end{equation}
hence, $\nu$ now parameterises differing levels of prior uncertainty.

The above description specifies a simple class of possible Bayesian analyses defined over a 2-dimensional space $\mathcal{X}$ where
\be
\mathcal{X} \;\;  \equiv \;\;  \left\{ x=(\nu,\epsilon) \; : \; \nu \in [0.3,2] \; {\rm and } \; \epsilon \in [0,1]
			\right\},
\ee
which is parameterised by the likelihood contamination and prior standard deviation parameters, $\epsilon$ and $\nu$ respectively, as summarised in Table~\ref{tab_toy_cm}.
\begin{table}
\begin{center}
\begin{tabular}{|c|c|c||c|c|}
\hline
 Inputs $x$ & Type of Input & Range & Outputs $f(x)$     \\
\hline
 $\nu$ & Prior standard deviation & $[0.3,2]$ & $\e{\theta| z,\nu,\epsilon}$    \\
 $\epsilon$ & Likelihood contamination & $[0,1]$ & $\SD{\theta| z,\nu,\epsilon}$   \\
\hline
\end{tabular}
\vspace{0.2cm}
\caption{\footnotesize{The inputs $x$ and outputs $f(x)$ of the example Bayesian model when represented as a computer model. The classes of inputs and outputs are also given along with the range of exploration of the inputs, defining the extent of the sensitivity analysis.}}\label{tab_toy_cm}
\end{center}
\end{table}

We now wish to explore the behaviour of attributes of the posterior $\pi(\theta| z,\nu,\epsilon)$ as a function of the inputs $\nu$ and $\epsilon$, and to investigate the corresponding robustness of these attributes and, hence, of the original analysis. We choose here to examine the posterior mean and standard deviation as these are usually of primary interest, but our approach could be applied to any set of posterior attributes. We define
\be 
f(x) \; = \; \left(\e{\theta| z,\nu,\epsilon},\SD{\theta| z,\nu,\epsilon}\right)
\ee
as the function to be explored, as also summarised in Table~\ref{tab_toy_cm}. Note that a perfunctory robust Bayesian analysis at this point may attempt to examine the range of possible values of the posterior attributes of interest, in this case the mean and standard deviation, that is achievable 
over $\mathcal{X}$. We wish to go further and to efficiently represent the posterior attributes for {\it any} choice of the inputs $\nu$ and $\epsilon$. This allows any expert to be able to extract their own Bayesian posterior attributes directly from our results, either corresponding to a particular specification represented as a single point in $\mathcal{X}$, or to a range of possible specifications represented by a subset of $\mathcal{X}$. 

As the specification is no longer in general conjugate, we construct a simple Metropolis-Hastings MCMC algorithm to allow evaluation of the posterior at any choice of inputs $\nu$ and $\epsilon$. As such a sampling algorithm is in some sense expensive (or would be for larger, more realistic models), we view the Bayesian updating process and its MCMC implementation as an expensive computer model, represented as the function $f(x)$, and employ computer model methodology in order to emulate and analysis the behaviour of $f(x)$. 


As the parameter of interest $\theta$ is non-negative, a Metropolis-Hastings MCMC algorithm was employed with a folded normal proposal distribution, 
\be
\theta^*|\theta_{t-1},\xi_\theta^2 \sim FN(\theta_{t-1},\xi_\theta^2),\\
\ee
where the folded normal has location parameter $\theta_{t-1}$, with the scale parameter fixed at $\xi_\theta^2=0.9$, which yielded reasonable acceptance rates between $0.30$ and $0.59$ for all evaluations of interest \citep{brooks2011handbook}. Note that the folded normal is still a symmetric proposal density, allowing for the usual simplification to the acceptance ratio. To avoid unnecessary complications in the description of our approach to this illustrative example, we minimised the MCMC sampling error and ensured convergence by running an excessively large number of steps. A total of 200000 steps were used, the initial condition 
$\theta= 0.5$ was chosen and a burn in of 100 steps assumed. 

An example of the posterior sample generated by the MCMC algorithm is given as the grey histogram in Figure~\ref{fig_conjcase} (right panel), for the initial conjugate case where $\epsilon=0$ and $\nu=1$. Also shown is the prior and true posterior distributions as the red and blue lines respectively. The prior mean 
$\mu_0$ and posterior mean $\e{\theta|z,\nu\!\!=\!\!1,\epsilon\!\!=\!\!0}$ are given as the vertical dashed red and blue lines respectively. This figure therefore represents a single point in the class of Bayesian updates $\mathcal{X}$ that we wish to emulate over, the specific point being 
$$f(x=(1,0))=\left(\e{\theta|z,\nu=1,\epsilon=0},\SD{\theta|z,\nu=1,\epsilon=0}\right).$$ 
For any point in $\mathcal{X}$ with $\epsilon > 0$, conjugacy is no longer true and the MCMC algorithm becomes vital. 

\subsubsection{Emulating the Bayesian analysis}\label{ssec_EmulBA}

In order to construct an emulator for the function $f(x)$ over $\mathcal{X}$, we now run the MCMC algorithm at a set of 35 design points $x_D$ over the two dimensional input space $\mathcal{X}$, using a lattice design. We check the convergence, the mixing plots and the autocorrelation plots for each of the 35 MCMC chains. As $\theta$ here is one dimensional, and as we employed a very large number of steps, our MCMC algorithm was unsurprisingly found to perform adequately across the whole input space (we discuss alternate MCMC strategies in section~\ref{ssec_future}). 

\begin{figure}
\begin{center}
\begin{tabular}{cc}
\hspace{-1.5cm} \includegraphics[scale=0.52,angle=0,viewport=-1 -1 430 385,clip]{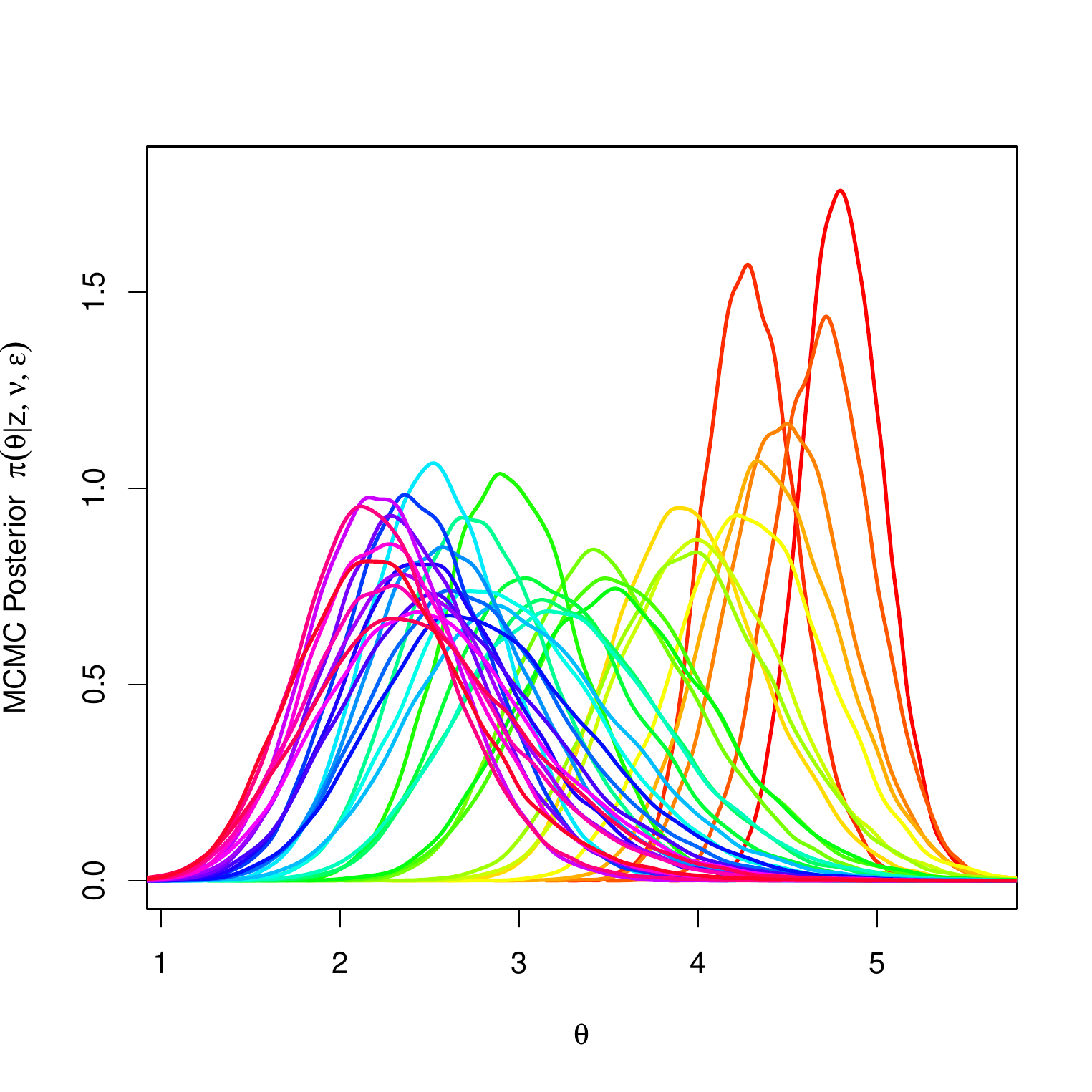}   &  
\hspace{-0.7cm}  \includegraphics[scale=0.47,angle=0,viewport=-1 -1 445 440,clip]{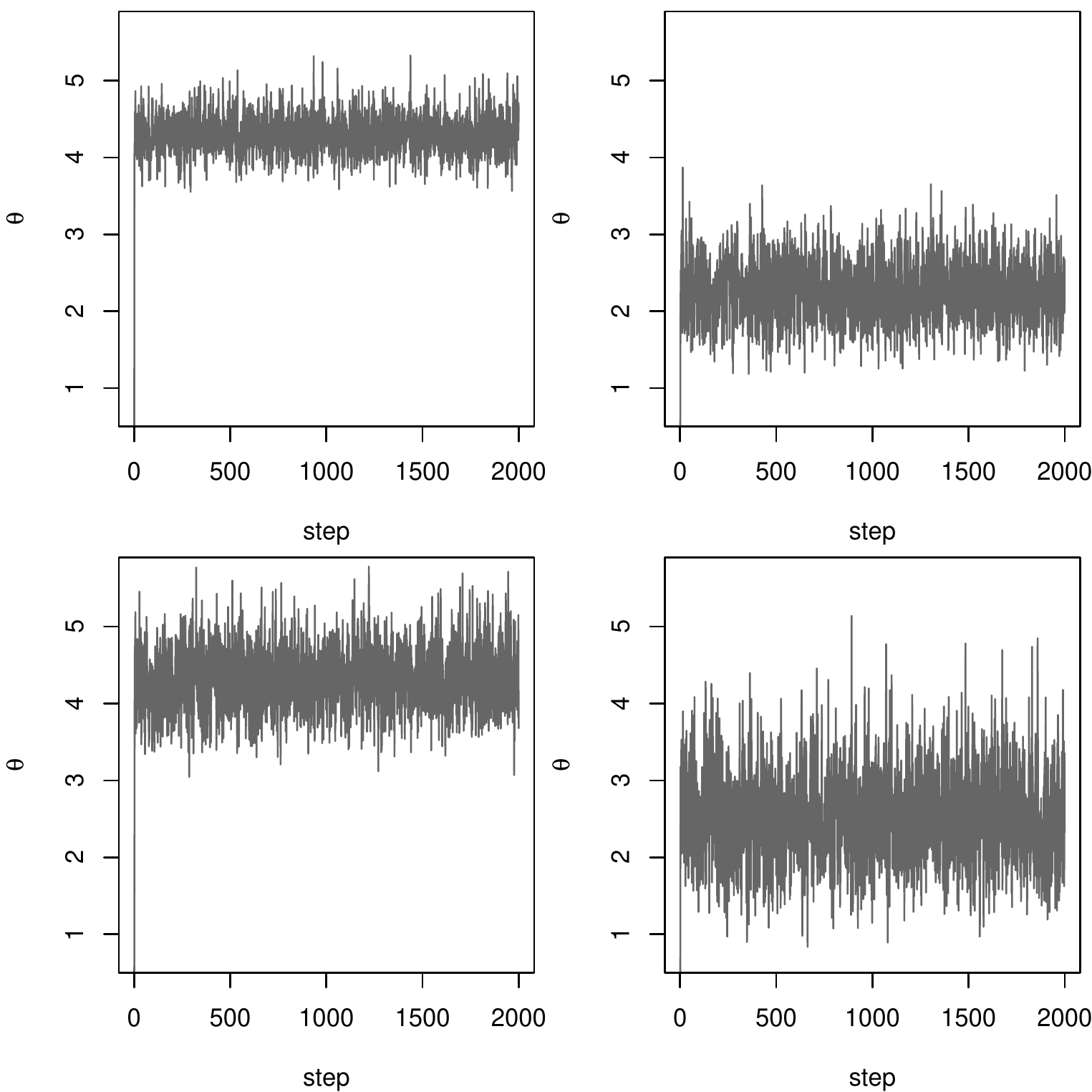} \\
\end{tabular}
\caption{\footnotesize{Left Panel: the estimated posterior density functions for the 35 MCMC runs performed across 
$\mathcal{X}$, coloured by their $\nu$ values, which display a reasonable range of posterior means and standard deviations. Right Panel: mixing plots for the four runs closest to the corners of the space $\mathcal{X}$, 
demonstrating excellent convergence and mixing, as expected.}}
\label{fig_35runs}
\end{center}
\end{figure}

Figure~\ref{fig_35runs} (left panel) shows the estimated posterior density functions for the 35 separate MCMC-based analyses performed across $\mathcal{X}$, which display a reasonable range of posterior means and standard deviations. This implies that the different choices within the analysis, as represented by $\mathcal{X}$, will lead to substantial differences in the Bayesian posterior. Note that this would likely preclude alternative strategies based on re-weighting one posterior sample to estimate other posterior 
attributes across $\mathcal{X}$, strategies that would likely become even weaker for more complex problems.
Figure~\ref{fig_35runs} (right panel) shows the mixing plots for the four runs closest to the corners of the space $\mathcal{X}$, and demonstrate convergence and excellent mixing, as expected. 

With such checks in place, we are now free to emulate the function $f(x)$ over the input space $\mathcal{X}$ using the methodology described in section~\ref{sec_BA_CCM}.
Specifically, we used a simple emulator construction sufficient for this example, with constant mean function $m(x)=m_0$, and covariance function $c(x,x')$ given by the Gaussian form of 
equation~(\ref{eq_cor_gauss}). We set the variance of the nugget equal to the mean of the MCMC sampling 
variance (a very small value), which was assumed constant across the input space. The inputs $x$ were scaled to 
have range $[-1,1]$ and a fixed correlation length of $\theta_{em}=0.6$ was used for both, following the 
arguments in \cite{Vernon10_CS,Vernon10_CS_rej} for choosing correlation lengths \emph{a priori}. Finally, the 
emulator variance parameter $\sigma^2_{em}$ was set equal to the variance of the 35 run outputs.

\begin{figure}
\begin{center}
\hspace{-0cm}
\begin{tabular}{cc}
\hspace{-1.2cm}\includegraphics[scale=0.48,angle=0,viewport=-1 -1 445 370,clip]{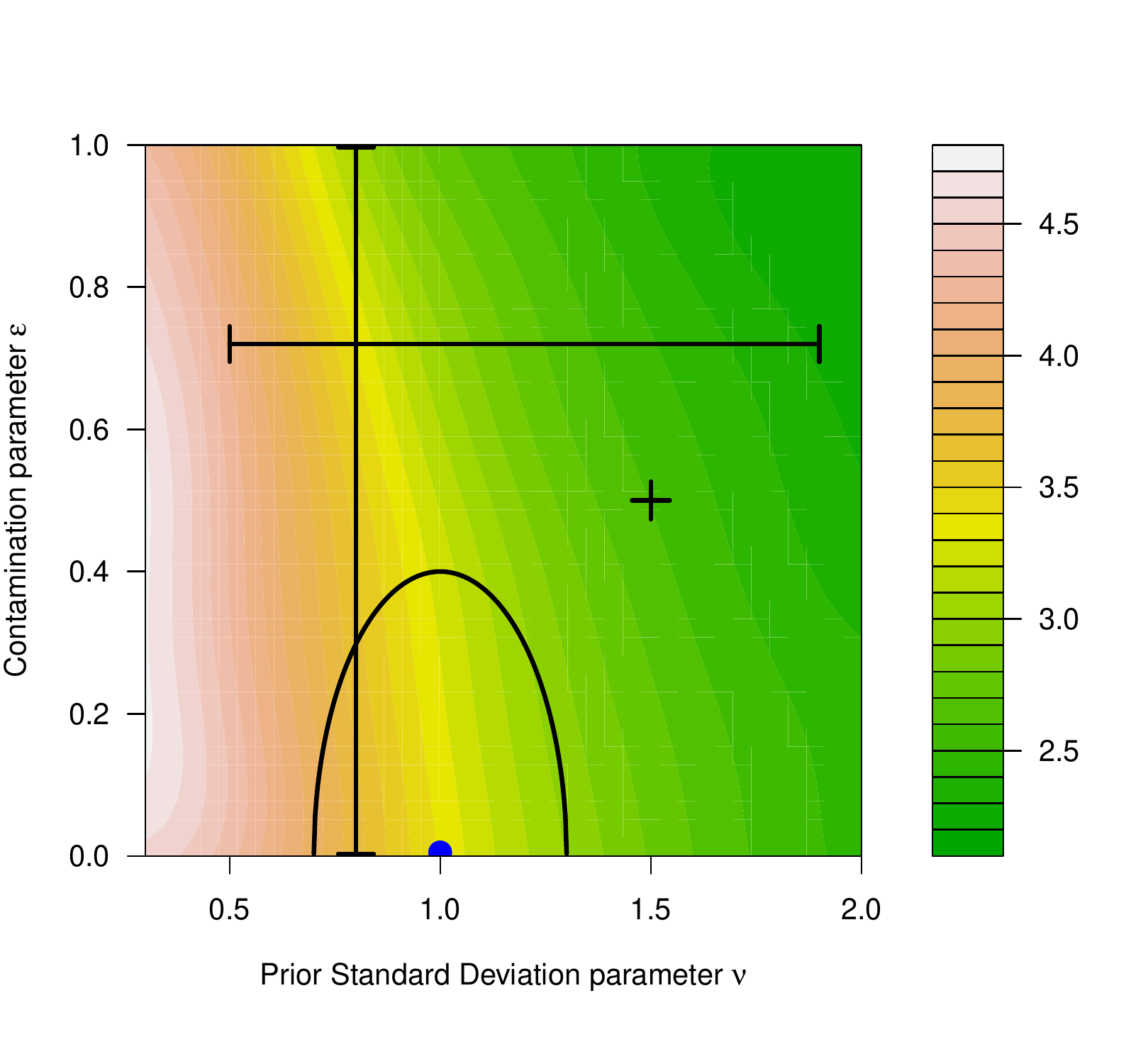} &
\includegraphics[scale=0.48,angle=0,viewport=-1 -1 445 370,clip]{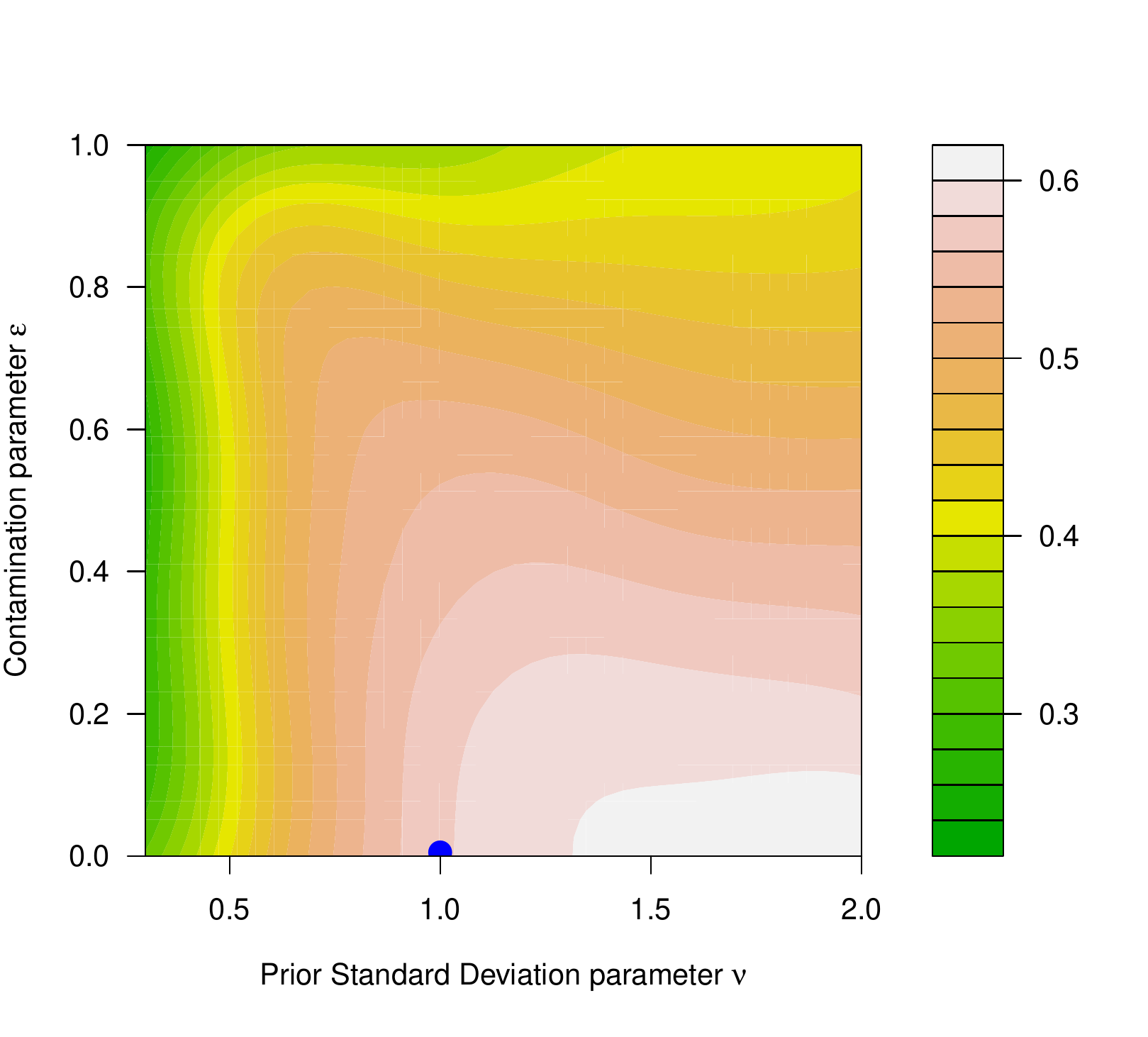} \vspace{-0.5cm} \\
\end{tabular}
\caption{\footnotesize{Left panel: the emulator expectation $\e{f_1(x)|f_1^{(s)}(x_D)}$ for the posterior mean $f_1(x) = \e{\theta|z,x}$ 
as a function of the inputs $x=(\nu,\epsilon)$, with the blue dot at the location of the original conjugate analysis, shown in Figure~\ref{fig_conjcase} (right panel). We see that the posterior mean is more sensitive to $\nu$ than to $\epsilon$, and that the original conjugate analysis is relatively robust to small departures from conjugacy, of the given form.  
Right panel: the emulator expectation $\e{f_2(x)|f_2^{(s)}(x_D)}$ for the posterior standard deviation $f_2(x) = \SD{\theta|z,x}$, 
as a function of the inputs $x=(\nu,\epsilon)$. Here, far more counterintuitive behaviour is displayed: for intermediate values of the contamination parameter $\epsilon$, an increase in the prior SD $\nu$ leads to a {\it decrease} in the posterior SD.}}
\label{fig_toy_emul}
\end{center}
\end{figure}

Figure~\ref{fig_toy_emul} (left panel) shows the emulator expectation $\e{f_1(x)|f_1^{(s)}(x_D)}$, given by equation~(\ref{eq_GPpostmean}), for the posterior mean 
$f_1(x) = \e{\theta|z,x}$ as a function of the inputs $x=(\nu,\epsilon)$ (we suppress the implicit conditioning on $m(.)$ and $c(.,.)$ as given in  equation~(\ref{eq_GPpost}) from here onward). The blue dot represents the original conjugate analysis 
where $x=(1,0)$, the output of which is shown in figure~\ref{fig_conjcase} (right panel). This plot instantly confirms several intuitive features 
about the class of Bayesian analyses, as well as providing clear quantitative statements in response to various robustness questions. 
We see that conditioning on $\epsilon$ and increasing $\nu$ always leads to a decrease in the posterior mean $\e{\theta|z,x}$, 
while conditioning on $\nu$ and increasing $\epsilon$ (and hence moving away from conjugacy) also decreases the mean. The experts may 
find it useful to know that moving away from conjugacy in this manner would lead to their posterior mean decreasing at most from 3.4 to approximately 2.75, as can be seen by drawing a vertical line above the blue dot, and that this mean is relatively insensitive to smaller 
likelihood contaminations of this form. A comparable lowering of the mean could also arise from choosing $\nu = 1.55$ instead of the original value of $\nu = 1$, showing 
that the analysis is far more sensitive to the prior standard deviation than to the likelihood contamination. For a careful interpretation, we should
also take account of the emulator variance $\var{f_1(x)|f_1(x_D)}$ and the corresponding credible intervals for $f_1(x)$ across $\mathcal{X}$, as 
is discussed for several example specifications in section~\ref{sec_ex_spec}.

Figure~\ref{fig_toy_emul} (right panel) shows the emulator expectation $\e{f_2(x)|f_2(x_D)}$ for the posterior standard deviation  
$f_2(x) = \SD{\theta|z,x}$ as a function of the inputs $x=(\nu,\epsilon)$, with the blue dot representing the original conjugate analysis. In contrast to the mean plot, this plot displays far more counterintuitive behaviour. Conditioning on $\nu$ and increasing $\epsilon$ has little effect for low $\nu$ and causes the SD to decrease monotonically for high $\nu$. When $\epsilon =0$ or 1, increasing $\nu$ 
leads to an increase in the posterior SD as expected. However, for intermediate values of the contamination $\epsilon$, 
there are regions of $\mathcal{X}$
for which the opposite is true: an increase in the prior SD $\nu$ leads to a {\it decrease} in the posterior SD. For example, 
a prior specification of $(\nu=0.8,\epsilon=0.72)$ has posterior SD = 0.50, but an increase in prior SD only to $(\nu=2,\epsilon=0.72)$ 
leads to a posterior SD = 0.46. So there are regions where being more certain a priori leads to one being comparatively less certain \emph{a posteriori}. Note that 
this is not due to an over interpretation of the SD which may be too simple a summary of complex distributions, as exactly the same effect is seen, for example, when examining the width of the corresponding HPD intervals. Nor is it an artefact of the emulation process, as has been checked by making further evaluations of the MCMC algorithm. 
Instead, this counterintuitive behaviour can be explained in terms of a wider, less restrictive prior allowing the Bayesian update to be influenced by  
a larger range of the contaminated likelihood, sections of which may favour posteriors with lower variance. Note that it would be impossible for a conjugate analysis to exhibit such behaviour.

Finding this non-trivial behaviour in a simple 1-dimensional case suggests that high-dimensional Bayesian analyses could easily exhibit similarly complex behaviour as we move away from conjugacy. The emulation methodology presented here is precisely designed to deal with high-dimensional cases of this form. Whether such complex behaviour was present would be difficult to discover without a careful global robustness/sensitivity analysis such as we propose here, which would be vital if the problem was deemed to be of high enough importance.

\subsubsection{Example specifications}\label{sec_ex_spec}

To further demonstrate the depth of analysis that is possible using Gaussian process emulation, we give the results of a small number of example specifications that could be provided by either single experts, or combinations of experts. We show 
that our analysis can give immediate and accurate answers in these cases, along with appropriate uncertainty statements that can be subsequently
used to decide if further runs of the MCMC algorithm are required, to achieve a desired level of accuracy. We imagine that the following four specifications have been made:
\vspace{0.3cm}
\begin{description}
\item[Case 1] An expert has precise prior beliefs $x_e$ corresponding to $\nu=1.5$ and $\epsilon=0.5$, but requests a local sensitivity analysis at this point.
\vspace{0.1cm}
\item[Case 2] The experts have a fixed prior variance but want to explore the full range of contamination: $\nu=0.8$, $0\le\epsilon\le1$.
\vspace{0.1cm}
\item[Case 3] The experts have a fixed level of contamination, but imprecise prior variance such that: $0.5\le\nu\le1.9$, $\epsilon=0.72$.
\vspace{0.1cm}
\item[Case 4] The experts wish to perform a robustness analysis over a half elliptical region around the original conjugate analysis ($\nu=1,\epsilon=0$) that satisfies \be
\frac{(\nu -1)^2}{0.3^2} + \frac{\epsilon^2}{0.4^2} \;\; < \;\; 1 \quad \quad {\rm and} \quad \quad \epsilon > 0.
\ee
\end{description}
\vspace{0.3cm}
These four cases are shown in Figure~\ref{fig_toy_emul} as the black cross and the black vertical, horizontal and curved lines respectively. 
The emulators derived in Section~\ref{ssec_EmulBA} can instantly provide the desired results for the four cases, as we now describe. 

Table~\ref{tab_case1} gives the emulator expectation (first row) for the posterior mean $f_1(x) = \e{\theta| z,\nu,\epsilon}$ and 
SD, $f_2(x) = \SD{\theta| z,\nu,\epsilon}$ (first and forth columns) evaluated at the point $x_e=(1.5,0.5)$, corresponding to the specification of case 1. As a local sensitivity analysis was 
requested, also given are the partial derivatives of $f_1(x)$ and $f_2(x)$ with respect to $\nu$ and $\epsilon$, at this point, calculated as described in section~\ref{sssec_appBA}. These show that 
$f_1(x)$ is sensitive to both $\nu$ and $\epsilon$ at $x_e$; however, $f_2(x)$ is relatively insensitive to changes in $\nu$. Most importantly, the second row 
of table~\ref{tab_case1} gives the uncertainties due to the emulation process corresponding to each of these quantities, in the form of the emulator standard deviations, found from equation~(\ref{eq_GPpostcor}). These can be used to determine if a desired level of accuracy has been achieved, or if further MCMC runs are required. 
\input NewPlots/Table1_altered.tex

\begin{figure}
\begin{center}
\begin{tabular}{cc}
\hspace{-1.1cm} \includegraphics[scale=0.52,angle=0,viewport=-1 15 370 340,clip]{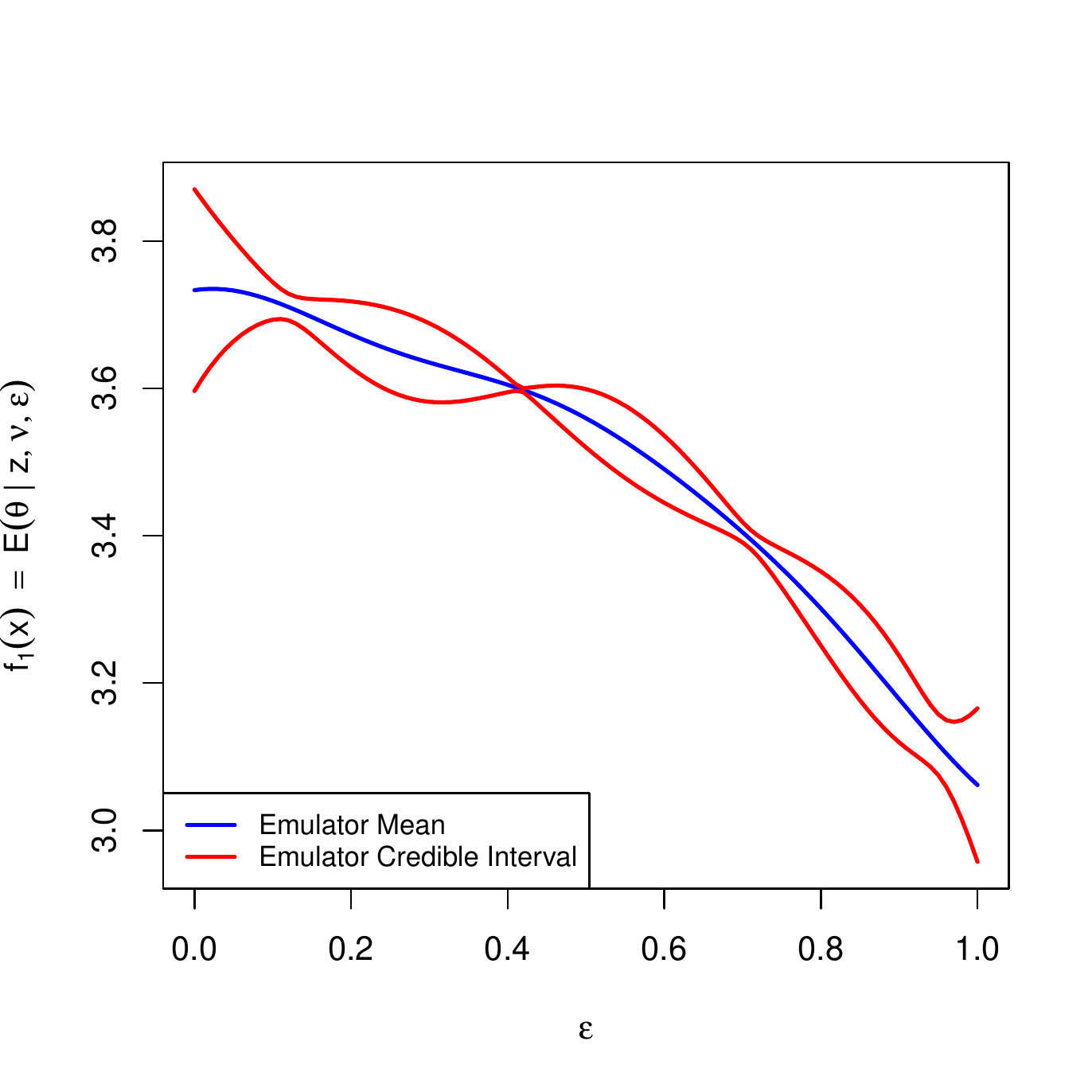}   &  
\hspace{-0.0cm}  \includegraphics[scale=0.52,angle=0,viewport=-1 15 370 340,clip]{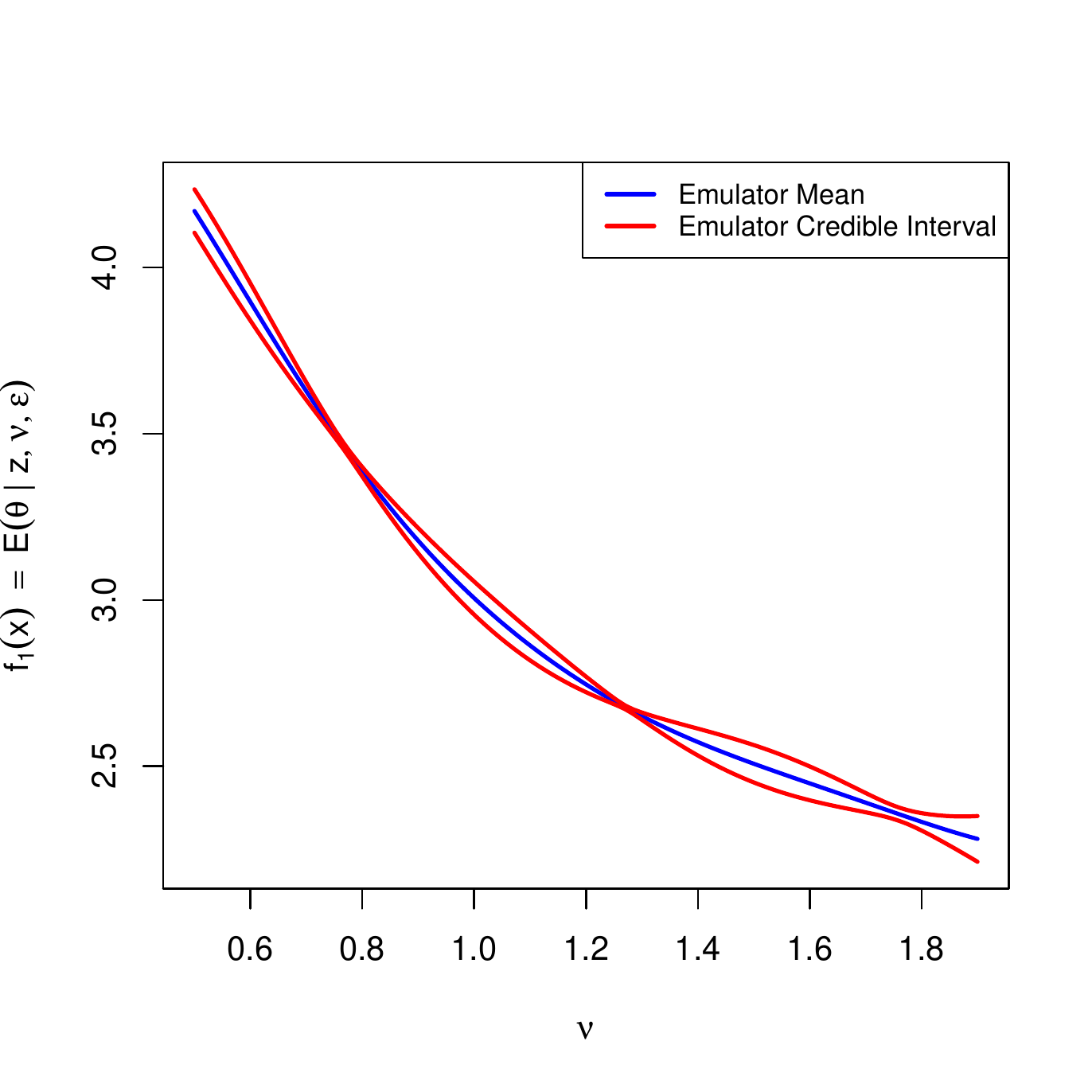} \\
\vspace{-0.0cm}
\hspace{-1.1cm} \includegraphics[scale=0.52,angle=0,viewport=-1 15 370 340,clip]{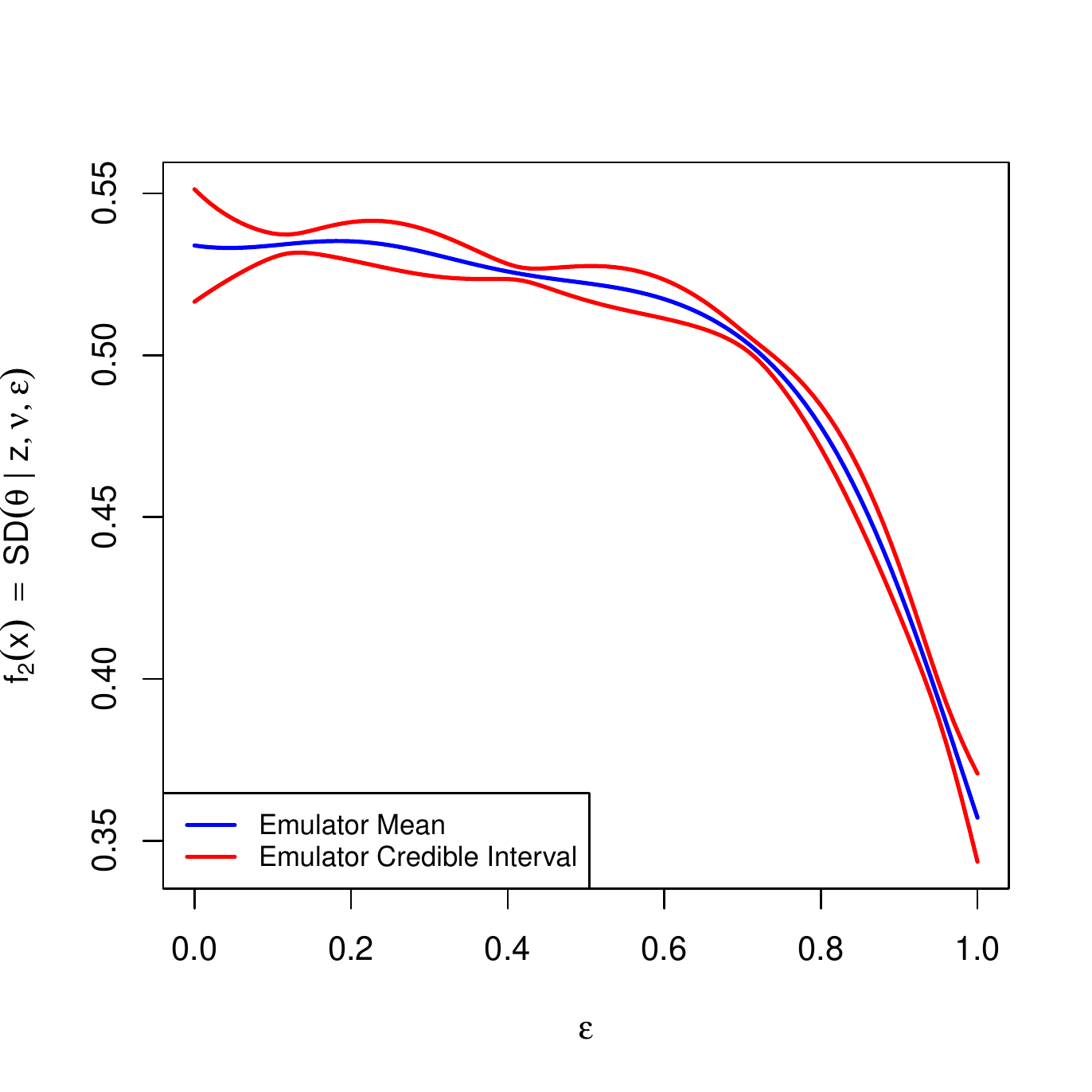}   &  
\hspace{-0.0cm}  \includegraphics[scale=0.52,angle=0,viewport=-1 15 370 340,clip]{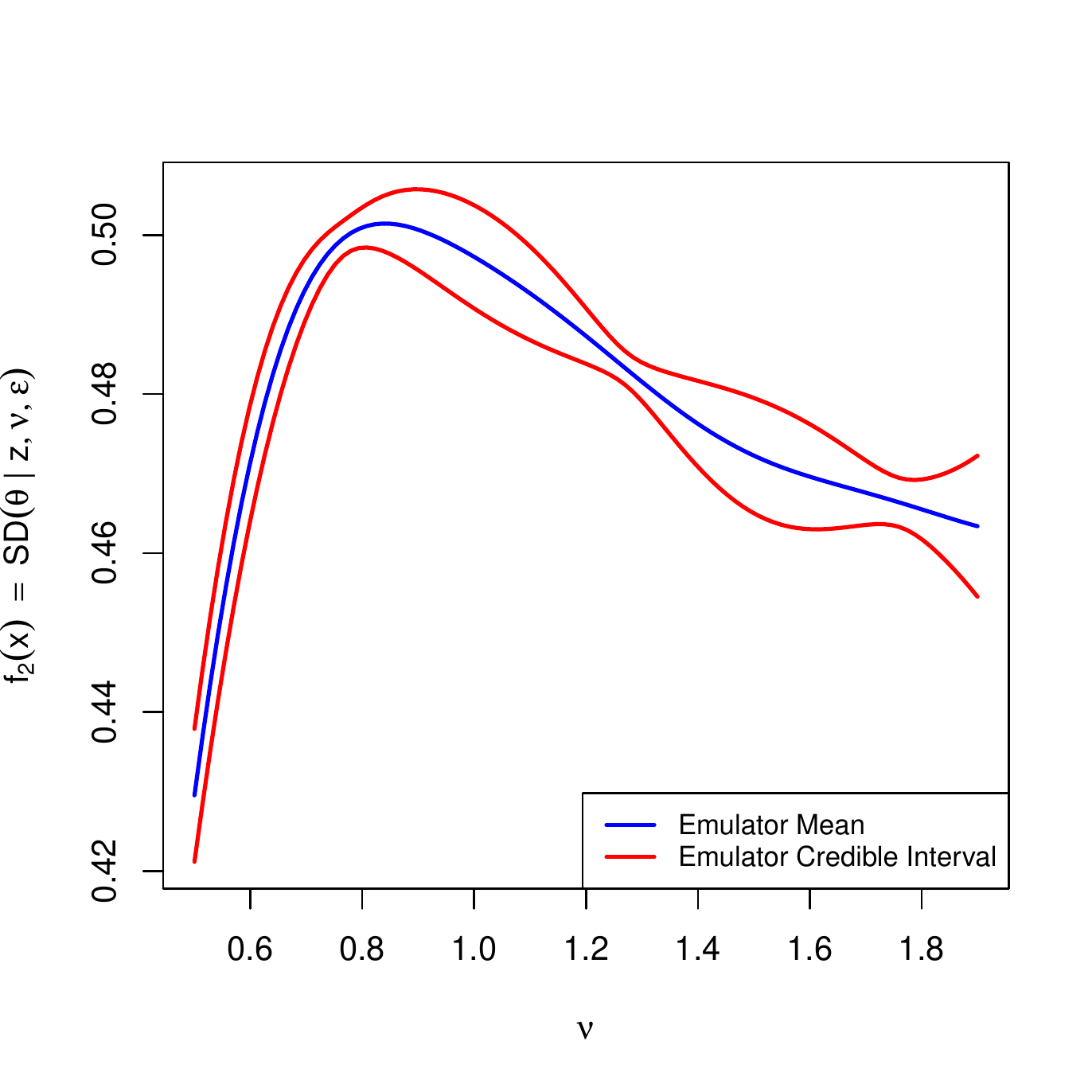} 
\end{tabular}
\caption{\footnotesize{Emulator expectations (blue lines) and 95\% credible intervals (red lines) for the expert specifications case 2 (left column) and case 3 (right column), with the results for the posterior means $\e{\theta| z,\nu,\epsilon}$ given by the top row, and the posterior  
standard deviations $\SD{\theta| z,\nu,\epsilon}$ by the bottom row.}}
\label{fig_emul_cases}
\end{center}
\end{figure}

Figure~\ref{fig_emul_cases} shows the results for the posterior mean $f_1(x)$ (top left panel) and posterior SD $f_2(x)$ (bottom left panel) for the specification given in case 2, where here the contamination parameter $\epsilon$ varies along the x-axis. The blue lines give the emulator expectations, 
and the red lines give a 95\% credible interval that represents the uncertainty due to the emulation process (and to a much smaller extent, due to the finite 
sample size of the MCMC draws). While both the posterior mean and SD appear to be monotonically decreasing with increasing $\epsilon$, the posterior SD sharply decreases for $\epsilon > 0.7$. This alerts the expert to the fact that careful thought may be required when specifying levels of contamination above 0.7.

Figure~\ref{fig_emul_cases} gives the corresponding plots (top right: posterior mean, bottom right: posterior SD) for the specification given in case 3, where the prior standard deviation parameter $\nu$ varies along the x-axis. We can see that the posterior mean has been emulated to a high degree of accuracy compared to 
its variation over this range. The posterior SD exhibits some of the counterintuitive behaviour discussed in Section~\ref{ssec_EmulBA}: once $\nu$ increases beyond
approximately 0.8, the posterior SD decreases as a function of increasing $\nu$. Here, the expert should be aware of both the counterintuitive behaviour and the sensitivity of the posterior to low values of $\nu$. 

In many situations, the experts may purely want a robust Bayesian analysis performed over their specified regions $\mathcal{X}_k$ say, that is the identification of the maximum and minimum of the posterior quantities of interest over $\mathcal{X}_k$. For the maximum, we would hence wish to evaluate $\e{\max_{x_e\in \mathcal{X}_k}{f(x_e)}}$, where the expectation is performed over the Gaussian process, 
however, unlike all examples up to this point, we do not have an analytic expression for this term, as the distribution of the maxima and minima of a Gaussian process is only known for a small number of specific correlation functions. However we can easily approximate these expressions using simulation as follows, being careful to respect the smoothness of $f(x)$ and hence the joint structure of the emulator over $\mathcal{X}_k$.
We define a large number of points $x^{(i)}_E$, $i=1,\dots, n_E$ spanning the specified region $\mathcal{X}_k$, and
simulate jointly from the emulator across $x^{(i)}_E$. 
Specifically, we use the joint posterior distribution 
over the vector $f(x_E)$ of length $n_E$, which is given by
\begin{equation}\label{eq_mvn_em}
f(x_E)|f^{(s)}(x_D),m(.),c(.,.) \;\; \sim \;\; N(m^*(x_E),\Sigma^*(x_E)),
\end{equation}
a direct consequence of equation~(\ref{eq_GPpost}),
where $\Sigma^*(x_E)$ is a covariance matrix of dimension $n_E$, with elements $\Sigma^*_{ij} = c^*(x^{(i)}_E,x^{(j)}_E)$. Equation~(\ref{eq_mvn_em}) can be used to efficiently simulate a large number $n_S$ of joint realisations from the posterior of the emulator. This provides,
\be
f^{(j)}(x^{(i)}_E), \quad {\rm with} \quad j=1,\dots,n_S \quad{\rm and}\quad i=1,\dots,n_E
\ee
From these we may extract $n_S$ maxima $M_j$ and minima $m_j$ and their corresponding means $\overline{M}$ and $\overline{m}$ respectively:
\ba
M_j \;=\; \max_i f^{(j)}(x^{(i)}_E) \; , \quad \quad && \quad \quad m_j \;=\; \min_i f^{(j)}(x^{(i)}_E)  \\
\overline{M} \;=\; \frac{1}{n_S} \sum_{j=1}^{n_S} M_j \; ,  \quad\quad && \quad\quad \overline{m} \;=\; \frac{1}{n_S} \sum_{j=1}^{n_S} m_j
\ea
where $\overline{M}$ and $\overline{m}$ are now estimates of $\e{\max_{x_e\in \mathcal{X}_k}{f(x_e)}}$ and $\e{\min_{x_e\in \mathcal{X}_k}{f(x_e)}}$ respectively. 

\begin{figure}
\begin{center}
\begin{tabular}{cc}
\hspace{-1.8cm} \includegraphics[scale=0.52,angle=0,viewport=-1 -1 412 380,clip]{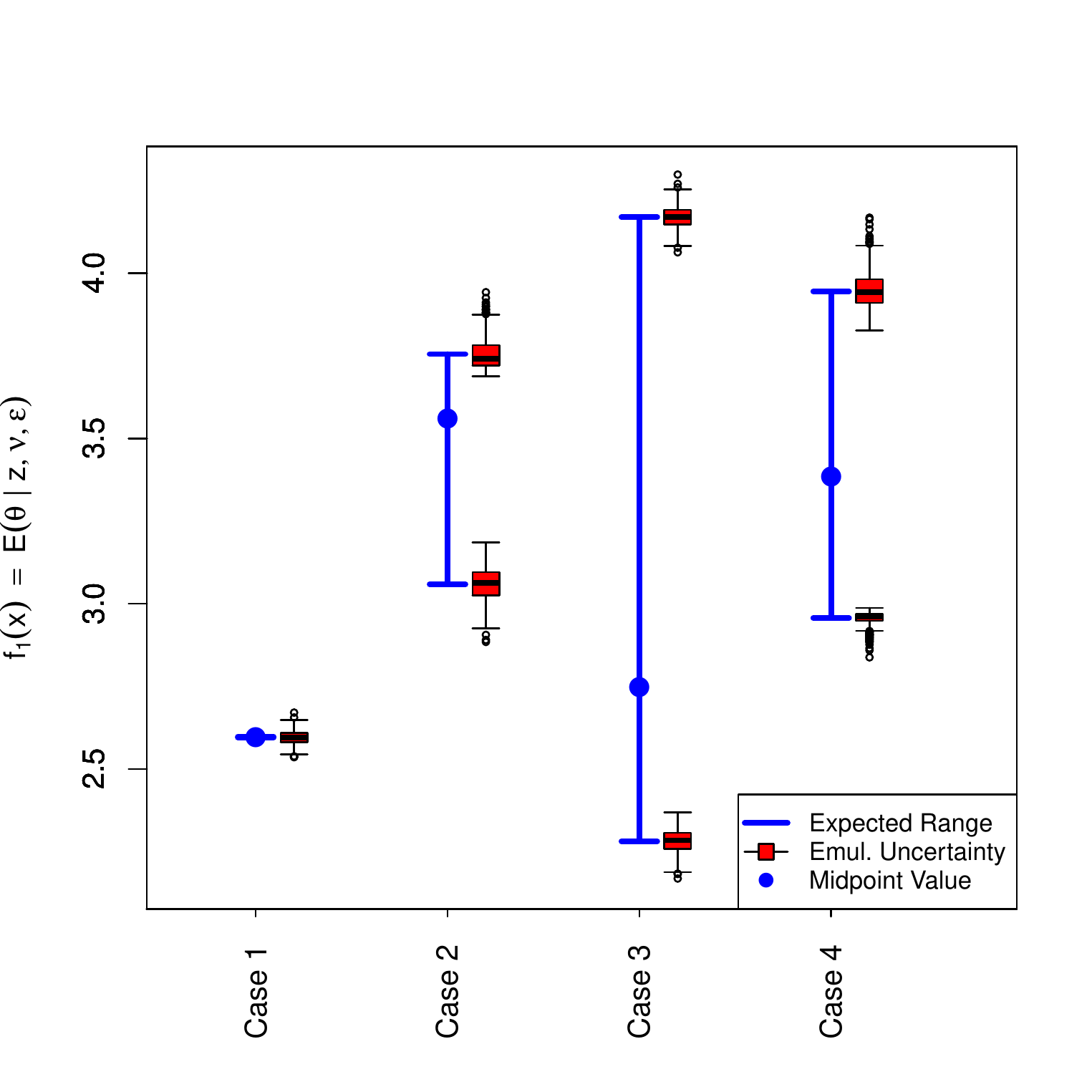}   &  
\hspace{-0.0cm}  \includegraphics[scale=0.52,angle=0,viewport=-1 -1 412 380,clip]{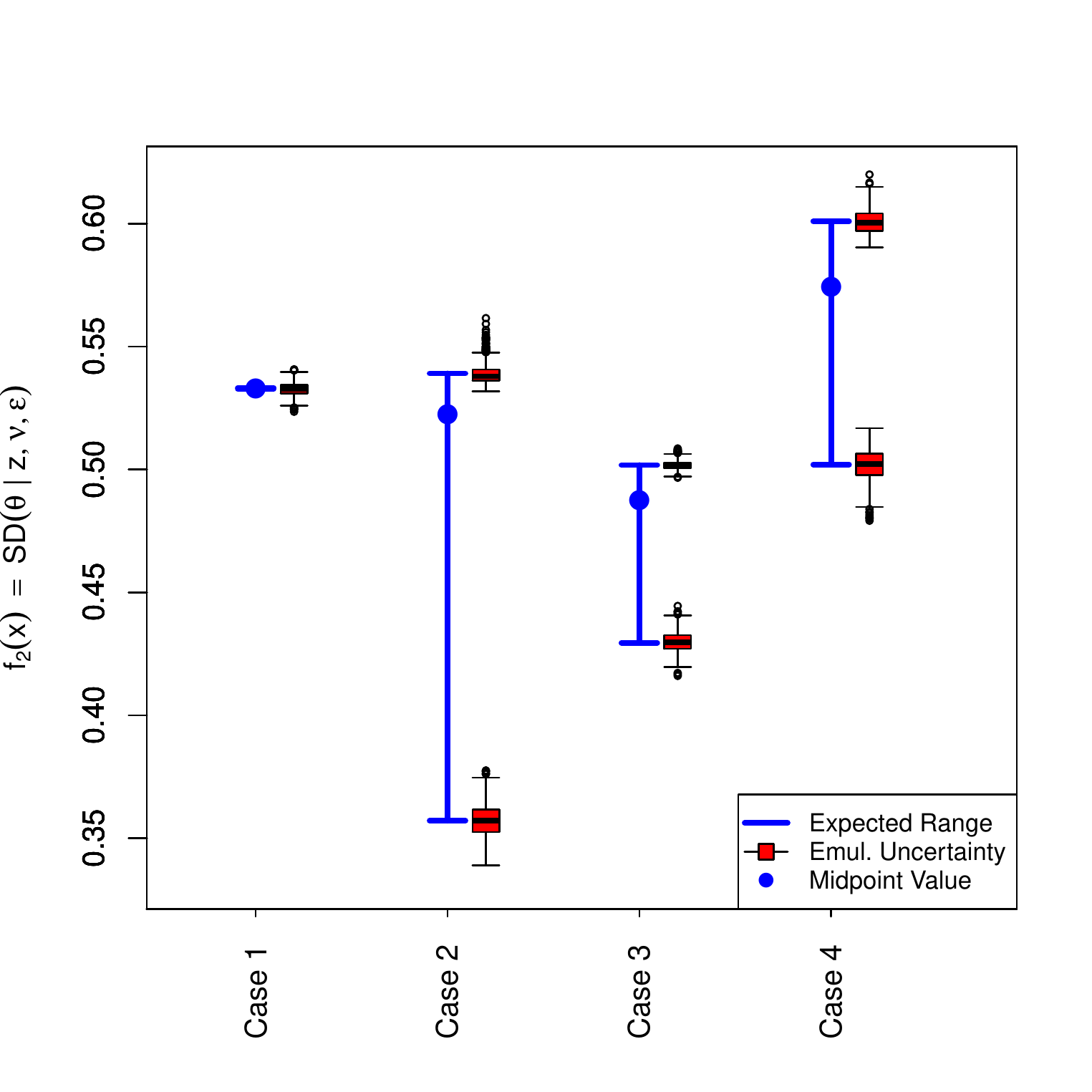} \\
\end{tabular}
\vspace{-0.7cm}
\caption{\footnotesize{Example output from the proposed analysis for the four example specifications given in Section~\ref{sec_ex_spec}. Left panel: the expected maximum and minimum of the posterior means $\e{\theta| z,\nu,\epsilon}$ are given by the top and bottom of the blue error bars. The uncertainty on these maximum and minimum estimates, due to the emulation process, is represented by the red box plots (based on 1000 realisations of the emulator), and could be reduced with further evaluations of the MCMC algorithm. The emulator expectation of the posterior mean at the midpoint of the imprecise specifications is given by the blue points. Right panel: the equivalent plot showing the possible ranges for the posterior standard deviations $\SD{\theta| z,\nu,\epsilon}$. Note the asymmetry of the box plots in some cases, due to the joint structure of the emulators.}}
\label{fig_case_sen_all}
\end{center}
\end{figure}

Figure~\ref{fig_case_sen_all} (left panel) shows the estimated expected maxima $\overline{M}$ and minima $\overline{m}$ and the intervening range of the posterior mean $f_1(x)$ as the blue error bars, where $\overline{M}$ and $\overline{m}$ are the top and bottom of the blue error bars respectively, for each of the four cases, as labelled on the $x$-axis. The uncertainty due to the emulation process regarding these maxima and minima is represented by the red boxplots, which are formed from $n_S=1000$ values of $M_j$ and $m_j$ respectively.
Note the resulting asymmetries in some of the boxplots: e.g. the maxima of case 2: 
this is due to the correlation structure of the underlying GP calculation, which still respects the smoothness of the Bayesian analysis as a function of $x$.
This can lead 
to accurate maximum and minimum estimates, even if the emulator uncertainty is high at individual input points.
The blue points show the expected posterior means evaluated at the midpoint of the specification region, for each case, which give an approximate idea of any non-linearity of the posterior mean's response.

\begin{figure}
\begin{center}
\begin{tabular}{cc}
\hspace{-1.6cm} \includegraphics[scale=0.44,angle=0,viewport=-1 -1 500 405,clip]{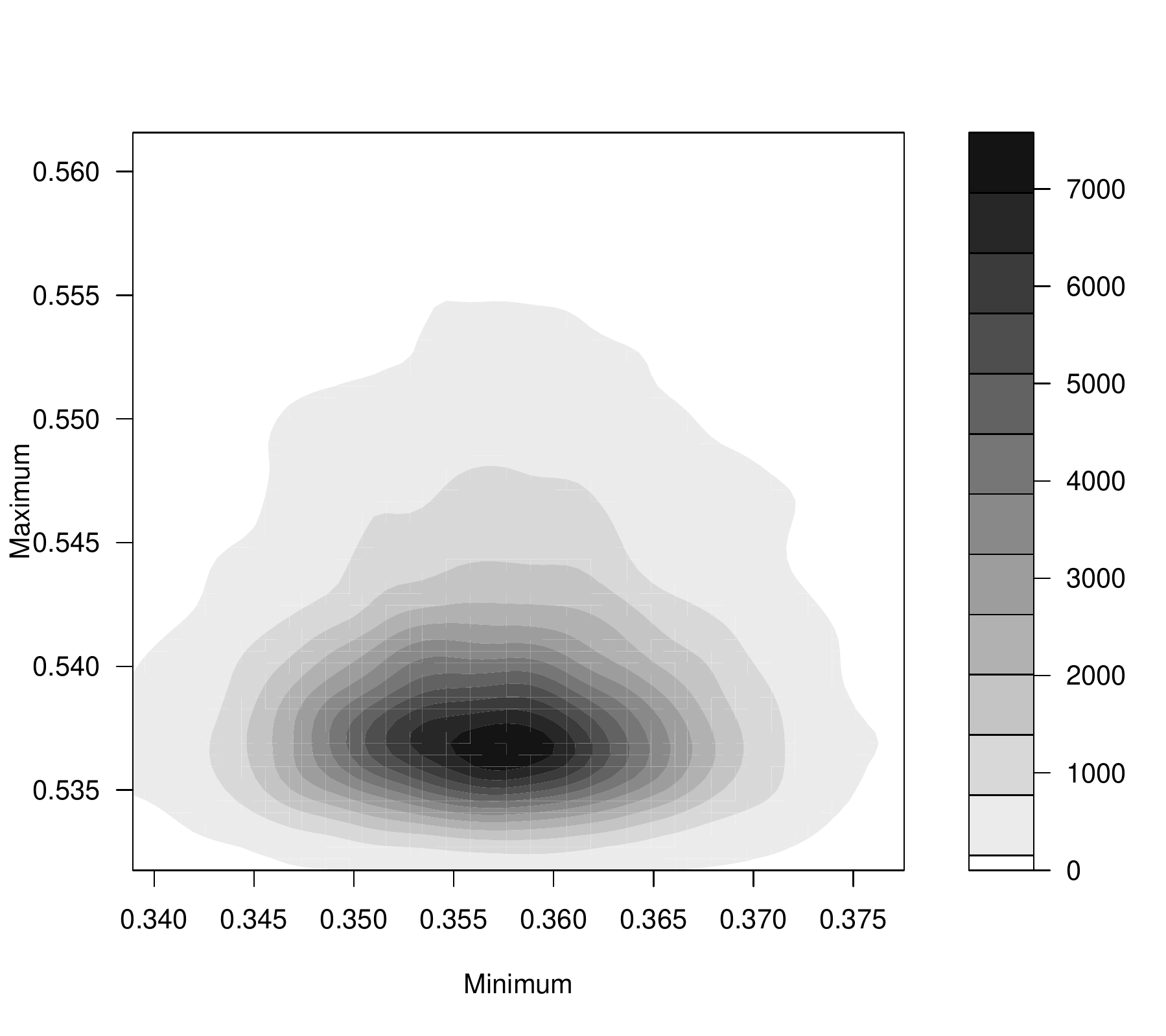}   &  
\hspace{-0.2cm}  \includegraphics[scale=0.44,angle=0,viewport=-1 -1 510 405,clip]{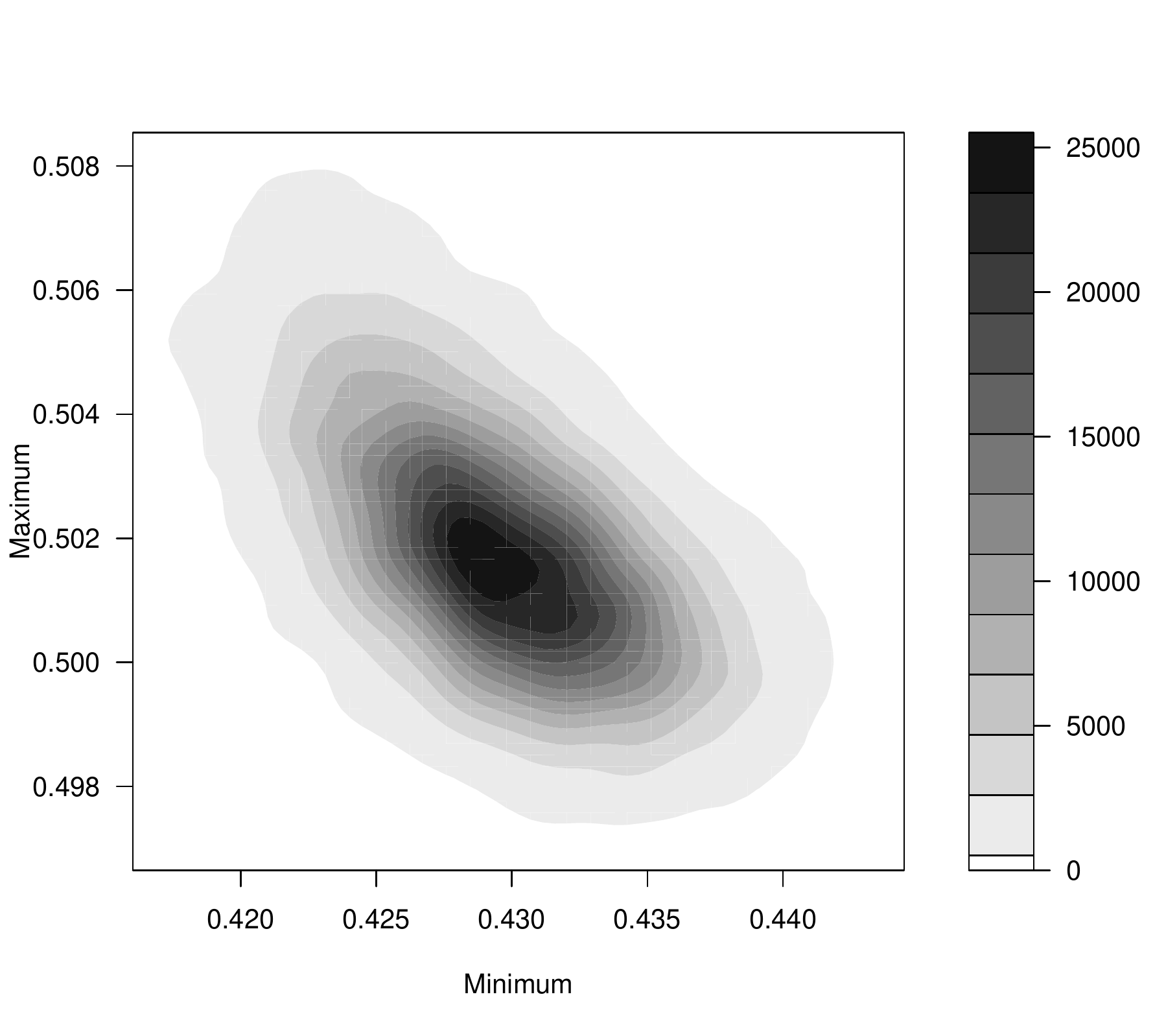}  \\
\end{tabular}
\vspace{-0.6cm}
\caption{\footnotesize{The empirically estimated joint emulator density of the maximum and minimum of the posterior standard deviations $\SD{\theta| z,\nu,\epsilon}$ for specification cases 2 (left panel) and 3 (right panel). The 1 dimensional marginals of these densities are given as the red box plots in Figure~\ref{fig_case_sen_all} (right panel). Note that in case 3 there is substantial negative correlation, a result of the maximum and minimum being located close together in input space, as can be seen in 
figure~\ref{fig_emul_cases} (bottom right panel), a fact incorporated in our analysis through the emulator's joint structure.}}
\label{fig_case_pdfs}
\end{center}
\end{figure}

The right-hand panel of Figure~\ref{fig_case_sen_all} shows the equivalent plot for the posterior standard deviations, $f_2(x)$. 
Once again, the emulator uncertainty, as represented by the red box-plots, shows
how much variation could be resolved by further MCMC runs. In both panels, for cases 2 to 4, we can see that we have captured the majority of the variation of the 
robust Bayesian analysis (as given by the blue error bars), and that the emulator uncertainty is small in comparison, so it is unlikely that we would wish to design more MCMC runs. However, as a result of the correlation structure of the updated emulator, given as $c^*(x,x')$ in equation~(\ref{eq_GPpostcor}), 
the uncertain maximum and minimum of the GP may be correlated, and even possess a complex joint structure. This can be seen in 
Figure~\ref{fig_case_pdfs}, which shows the empirical joint emulator density function of the maximum and minimum of the posterior standard deviations $\SD{\theta| z,\nu,\epsilon}$ for specification cases 2 (left panel) and 3 (right panel), based on 1000 realisations of $M_j$ and $m_j$. The 1 dimensional marginals of these densities are given as the red box plots in Figure~\ref{fig_case_sen_all} (right panel). Note that in case 3 there is substantial negative correlation, a result of the maximum and minimum being located close together in input space, as can be seen in 
Figure~\ref{fig_emul_cases} (bottom right panel), which would result in a larger uncertainty in the range of the posterior standard deviation for this case.

We now imagine that there is an important decision criteria that demands an alternative action if the posterior mean $f_1(x) < 2.6$ and the
posterior standard deviation $f_2(x)<0.47$ say. Experts in cases 1, 2 and 4 can rule out the alternative action immediately, as our analysis has confirmed 
that despite the imprecision in their specifications, their posteriors will not be close to the critical region. In case 3, these criteria are indeed possible, and the expert now knows that they need to think carefully about their original specification, particularly for the higher values of $v$ where the critical region lies, as is confirmed by Figure~\ref{fig_emul_cases} (right column). Should we need to do further exploration of the Bayesian analysis, in order to reduce the uncertainty about the location of such a critical region, we would perform additional waves of MCMC runs, using the well developed history matching methodology \citep[see][]{Vernon10_CS,Vernon10_CS_rej}, as is discussed further in Section~\ref{ssec_future}.


\section{Application to a Bayesian analysis of river flow}\label{sec_riverflowiv}

\subsection{Extension of a conjugate analysis}\label{ssec_ext_con_ana}

\cite{vicens1} give an account of a conjugate Bayesian analysis of annual streamflows of the Pemigewasset River at a measuring point at Plymouth, New Hampshire, USA. The data were the recorded flows in $\mbox{ft}^3/\mbox{s}$ over the 60-year period of 1904-1960 \citep{USGS1}. In their calculations, they assumed that the annual streamflows were identically and independently distributed as
\begin{eqnarray*}
z_i|\mu,\sigma^2 &\sim& \mbox{N}\left(\mu,\sigma^2 \right), \quad\quad i=1,\dots,60,
\end{eqnarray*}
where $\mu$ and $\sigma^2$ were parameters that they wished to learn about. In order to have a conjugate analysis, the following prior specification for $\mu$ and $\sigma$ was made:
\begin{eqnarray*}
\mu|\sigma^2 &\sim& \mbox{N}\left\{\mu_0,\left(\frac{\sigma}{n_0}\right)^2\right\},\\
\sigma^2 &\sim& \mbox{Inv-Ga}(\alpha,\beta),
\end{eqnarray*}
where  $\mu_0$, $n_0$, $\alpha$ and $\beta$ are hyperparameters that were specified. 

We embed their analysis within a more general structure as follows. 
Because the data can be naturally thought of as a time series, the following simple extension can be made to the assumed data generating process:
\begin{eqnarray*}
z_i-\phi(z_{i-1}-\mu)|\mu,\sigma^2,\phi &\sim& \mbox{N}\left(\mu,\sigma^2 \right), \quad\quad i=2,\dots,60,\\
z_1|\mu,\sigma^2 ~~~&\sim& \mbox{N}\left(\mu,\sigma^2 \right),
\end{eqnarray*}
where $\phi$ is a correlation parameter that could be fixed or we may be uncertain about. Because we are aiming to demonstrate just some of the utility of our approach and we have limited knowledge of the problem in hand, we will also investigate the following extension of the prior specification of Vicens \emph{et al.} (1975):
\begin{eqnarray*}
\mu|\sigma^2 &\sim& (1-\epsilon)\mbox{N}_Q\left(Q_1,Q_3\right)+ \epsilon~\mbox{C}_Q\left(Q_1,Q_3\right),\\
\sigma^2 &\sim& \mbox{Inv-Ga}(\alpha,\beta),
\end{eqnarray*}
where $Q_1$ and $Q_3$ denote the lower and upper quartiles respectively and $\mbox{N}_Q$ and $\mbox{C}_Q$ are normal and Cauchy distributions that are parameterised using the lower and upper quartiles derived from
$$\mbox{N}\left\{\mu_0,\left(\frac{\sigma}{n_0}\right)^2\right\}.$$
In order to complete this extended specification, we need to assign values to $\mu_0$, $n_0$, $\alpha$, $\beta$, $\phi$ and $\epsilon$.

When we use this prior specification with $\epsilon\neq 0$, we lose conjugacy and we need some numerical technique to derive the posterior distribution. For this application, we use a Metropolis-Hastings algorithm with proposal distributions:
\begin{eqnarray*}
\mu^*|\mu_{t-1},\xi_\mu^2 &\sim& N(\mu_{t-1},\xi_\mu^2),\\
\sigma^{2*}|\sigma_{t-1}^2,\xi_\sigma^2 &\sim& N(\sigma_{t-1}^2,\xi_\sigma^2).
\end{eqnarray*}
We use an adaptive algorithm to choose $\xi_\mu^2$ and $\xi_\sigma^2$, and we use diagnostics to ensure the convergence of the Markov chains for each set of hyperparameters as in Section~\ref{ssec_EmulBA}.

\subsection{Emulation of the Bayesian analysis}\label{sec_em_ba_rain}

We take as inputs to the computer model the specified parameters $x=  (\mu_0,n_0,\alpha,\beta,\phi,\epsilon)$. We take as outputs $f(x)$ the posterior mean and variances of both $\mu$ and $\sigma^2$. 
The inputs and outputs are listed in Table~\ref{tab_full_cm} along with the ranges we decided to explore the analysis over that define the region $\mathcal{X}$.

\begin{table}[ht]
\begin{center}
\begin{tabular}{|c|c|c||c|c|}
\hline
Inputs $x$ & Type of Input & Range & Outputs $f(x)$     \\
\hline
$ \mu_0$ &  Prior hyperparameter & [500, 2000] & $\e{\mu|z}$  \\ 
$n_0$ & Prior hyperparameter & [0.5, 30] &$\var{\mu|z}$ \\
$\alpha$ & Prior hyperparameter & [100, 500] &  $\e{\sigma^2|z}$ \\
$\beta$ & Prior hyperparameter &  [0, 30] & $\var{\sigma^2|z}$ \\
$\phi$ & Autocorrelation parameter & [-0.2, 0.5]  &   \\
$\epsilon$ & Prior contamination & [0,1] &  \\
\hline
\end{tabular}
\caption{\footnotesize{The inputs $x$ and outputs $f(x)$ of the extended Bayesian analysis of river flow when represented as a computer model. The ranges for the inputs are also given to define the extent of the sensitivity analysis over $\mathcal{X}$.}}\label{tab_full_cm}
\end{center}
\end{table}

We create a 100-point design by creating a 99 point maximin Latin hypercube over the six dimensional hypercube $\mathcal{X}$ given by the ranges in Table~\ref{tab_full_cm} and adding a single input corresponding to the particular conjugate analysis carried out in \cite{vicens1}. The parameters for the conjugate analysis were: $\mu_0 = 1,333$, $n_0=1$, $\alpha=6.5$, $\beta=402,057.5$, $\phi=0$ and $\epsilon=0$. We created a training set for our emulator by running the MCMC algorithm for each of the parameters and recording the four posterior moments of interest. The emulator was built using a Mat\'{e}rn correlation function, a linear mean function and an extra variance term to capture variability in the MCMC estimation process as described in Section~\ref{ssec_EmulBA}. We also checked the performance of the emulator using the diagnostic tools of \cite{bastos1}, and we found that the uncertainty caused by employing an emulator was generally two orders of magnitude smaller than the range of different values we observed for each of the four outputs of interest.

\begin{figure}
\begin{center}
\begin{tabular}{ll}
\hspace{-1.2cm} \includegraphics[scale=0.42,angle=0,viewport=0 15 490 460,clip]{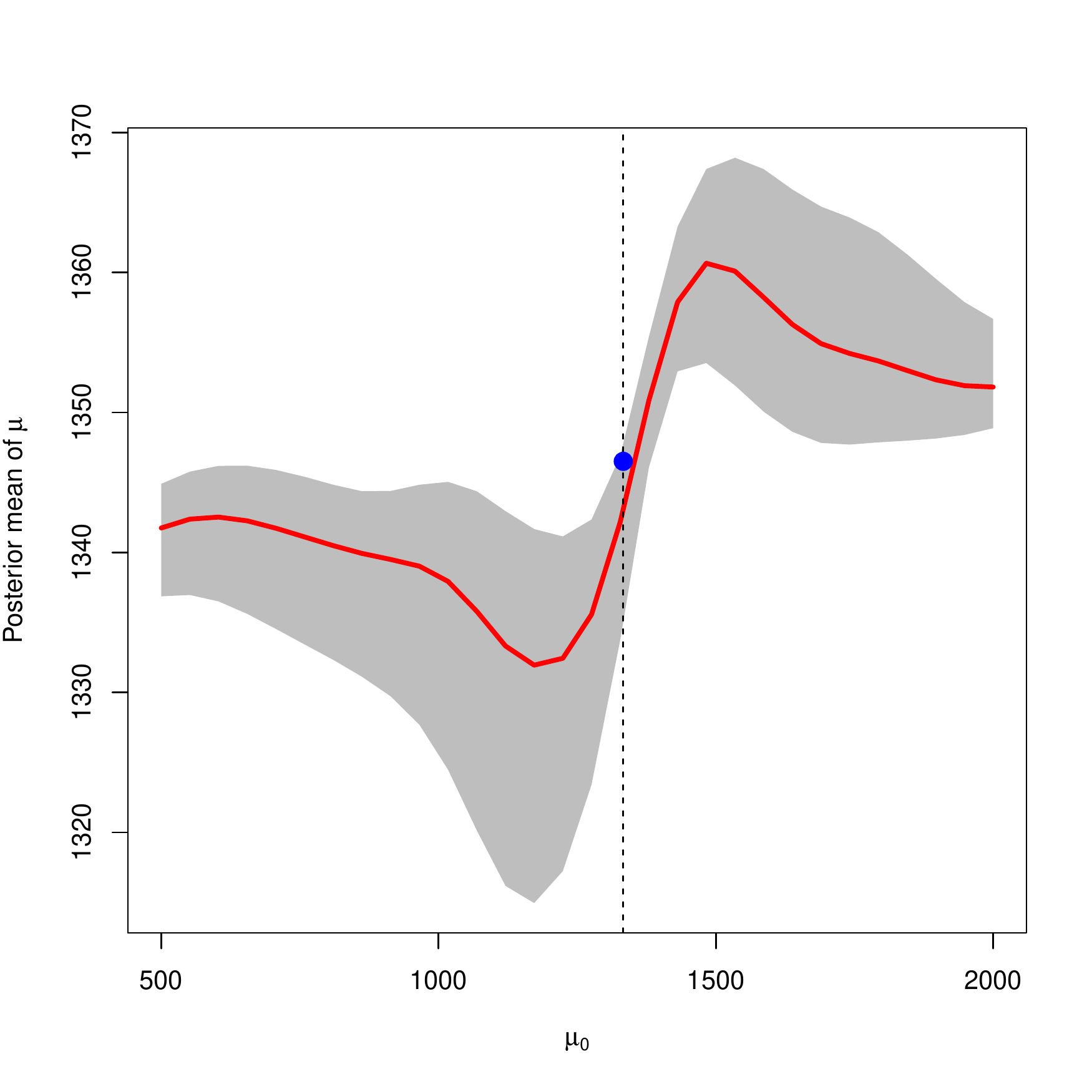} & \hspace{-0.5cm}
\includegraphics[scale=0.42,angle=0,viewport=0 15 480 460,clip]{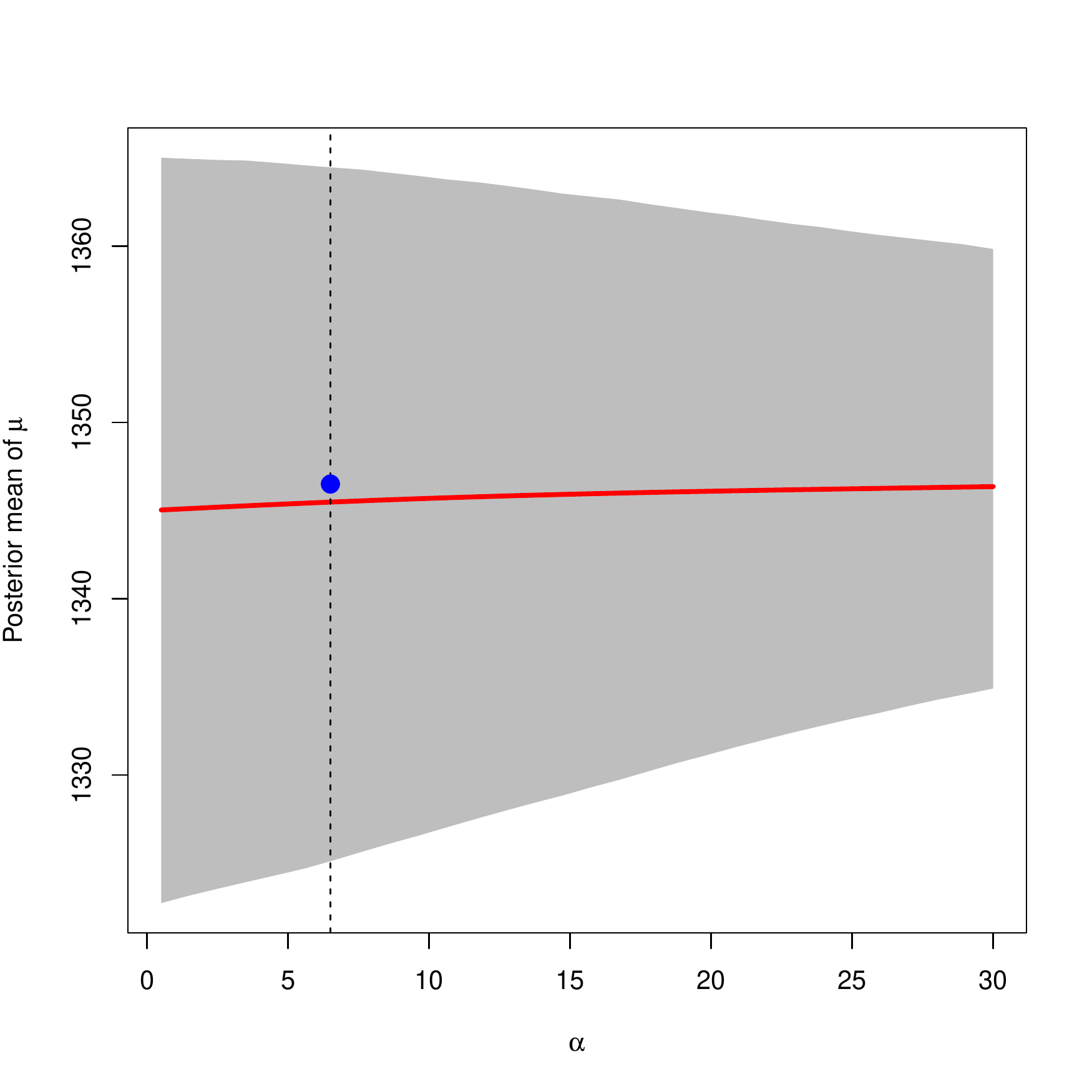} \\ 
\hspace{-1.2cm} \includegraphics[scale=0.42,angle=0,viewport=0 15 480 460,clip]{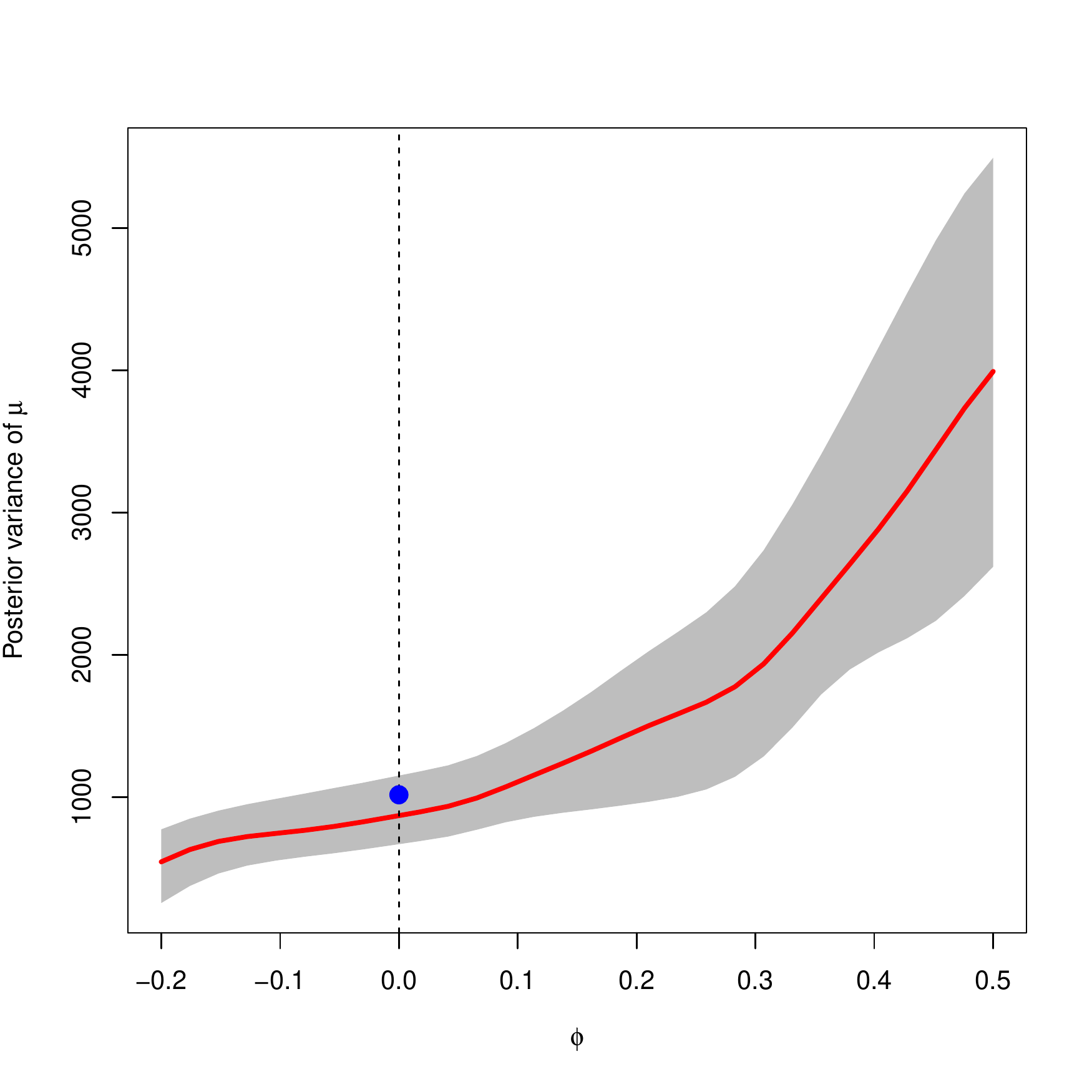} & \hspace{-0.5cm}
\includegraphics[scale=0.42,angle=0,viewport=0 15 480 460,clip]{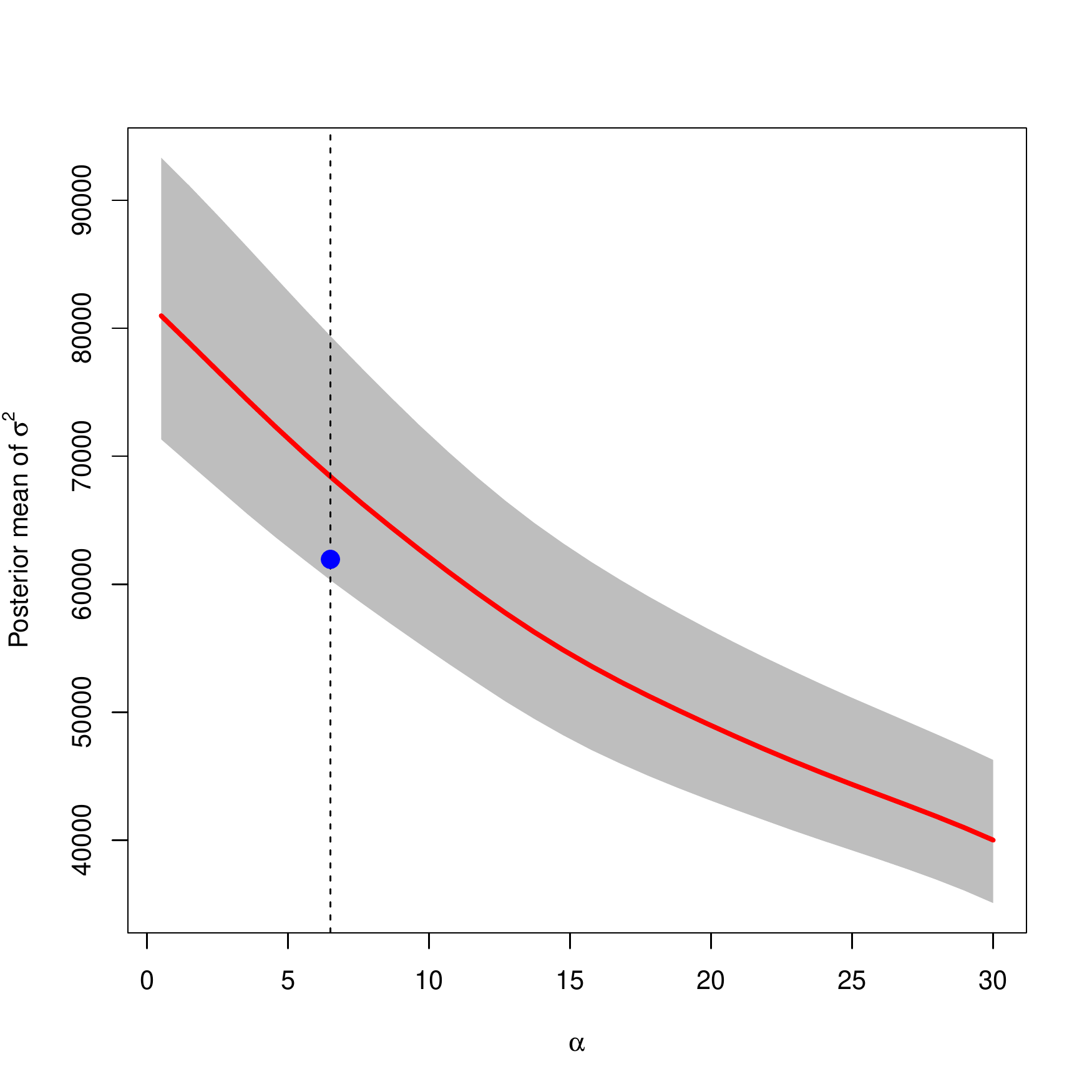} \\
\end{tabular}
\caption{\footnotesize{Main effects plots, showing the average effect of inputs $\mu_0$, $\alpha$ and $\phi$ on various outputs (red line). The grey envelope represents the range of possible output values (due to varying the other five inputs), conditioned on the given input. The blue points give the outputs corresponding to the prior specification for the parameters made by \cite{vicens1}.}}
\label{fig_full_me}
\end{center}
\end{figure}

Figure~\ref{fig_full_me} shows the effect of changing some of the parameters for three of the outputs of interest. The red line in each plot gives the average value for the output named on the y-axis conditional on the fixed value of the input from the x-axis. For these sensitivity analysis plots, we assume uniform distributions over all the ranges given in Table~\ref{tab_full_cm}. The grey regions on the plot show a 90\% credible interval for the different outputs conditional on the fixed input value and can be thought of being illustrative of plausible values for the output given the fixed input value. The top-right plot of Figure~\ref{fig_full_me} shows that as we vary $\alpha$ the posterior mean of $\mu$ will on average stay at 1347 $\mbox{ft}^3/\mbox{s}$, but the plausible range of values shrinks slightly as we increase $\alpha$. We can of course create such plots for each input-output combination, and the four shown in Figure~\ref{fig_full_me} are the most interesting for this example in that the ranges and mean change over the range of the input. The top-left plot of Figure~\ref{fig_full_me} shows the potentially unexpected effect that changing $\mu_0$ has on the posterior mean of $\mu$ in this analysis: relatively small deviations from the original specification of $\mu_0$ can have a large effect on the posterior mean of $\mu$. This information is obviously useful to any interested in the robustness and sensitivity of this hyperparameter in this analysis. The fact that the posterior mean is, on average, stable for values of $\mu_0$ below 1,000 and above 1,700 would be of interest to scientists who have prior beliefs that accord with one of those possibilities in that they will know that relatively little effort should be spent on eliciting their beliefs about $\mu_0$ precisely. This behaviour is due to the Cauchy part of the prior distribution dominating the Bayesian update, when there is modest prior-data conflict. We must of course interpret such plots with caution, and may choose to further investigate interesting features with a second wave of MCMC runs, as is discussion in section~\ref{ssec_future}.

\vspace{-0.5cm}
\hspace{0cm}

The variance contributions of each of the inputs to each of the outputs of interest are calculated to show the influence of each input using the probabilistic sensitivity analysis method of Oakley and O'Hagan (2004) (again, using uniform distributions over the ranges in Table~\ref{tab_full_cm}). The results are given in Tables~\ref{tab_mu} and \ref{tab_sig}. In the tables, the main-effect index column shows the percentage of variance in the output that is due to the corresponding input alone, and the total-effect index is the percentage of variance that is due to the corresponding input and all of the higher-order interactions it is involved in \citep{saltelli1}. Immediately, from the tables, we can see which parameters have most impact on the different outputs of interest: for instance, we can see that (over the ranges specified) $\mu_0$ is accounting for the majority of the variation in the output E$(\mu|z)$ as expected, but $\mu_0$ is having no discernible effect on any other output of the analysis. The impact of the contamination parameter $\epsilon$ across all the analysis outputs can also be seen to be relatively small, which may be of interest to any person who questioned the choice of the normal prior in the original paper. From Table~\ref{tab_mu}, we can also see that the input $\phi$ is having an effect on   E$(\mu|z)$, but only in interaction with the other input parameters (most probably $\mu_0$). Given this information, we may want to investigate the changes in E$(\mu|z)$ when we jointly manipulate  $\mu_0$ and $\phi$.
\begin{table}[ht]
\begin{center}
\begin{tabular}{c|c|c|c|c|}\cline{2-5}
&\multicolumn{2} {c|}{E$(\mu|z)$}&\multicolumn{2} {c|}{Var$(\mu|z)$}\\\cline{2-5}
 & Main-effect& Total-effect & Main-effect & Total-effect  \\ 
 & index (\%) & index (\%) &  index (\%) & index (\%) \\ \hline
\multicolumn{1} {|c|}{$\mu_0$} & 71 & 99&  ~0 & ~0 \\ 
\multicolumn{1} {|c|}{$n_0$} & ~0 & ~4 &  ~0 & ~4 \\ 
\multicolumn{1} {|c|}{$\alpha$} & ~0 & ~2 &  11 & 21 \\ 
\multicolumn{1} {|c|}{$\beta$} & ~0 & ~0 &  ~1 & ~5 \\ 
\multicolumn{1} {|c|}{$\phi$} & ~1 & 24 & 75 & 85 \\ 
\multicolumn{1} {|c|}{$\epsilon$} & ~0 & ~1 & ~0 & ~5 \\  \hline
\end{tabular}
\caption{\footnotesize{Variance-based sensitivity indices for the $\mu$ outputs}}\label{tab_mu}
\end{center}
\end{table}
\vspace{-0.5cm}
\begin{table}[ht]
\begin{center}
\begin{tabular}{c|c|c|c|c|}\cline{2-5}
&\multicolumn{2} {c|}{E$(\sigma^2|z)$}&\multicolumn{2} {c|}{Var$(\sigma^2|z)$}\\\cline{2-5}
 & Main-effect& Total-effect & Main-effect & Total-effect  \\ 
 & index (\%) & index (\%) &  index (\%) & index (\%) \\ \hline
\multicolumn{1} {|c|}{$\mu_0$} & ~0 & ~0&  ~0 & ~0 \\ 
\multicolumn{1} {|c|}{$n_0$} & ~0 & ~0 &  ~0 & ~1 \\ 
\multicolumn{1} {|c|}{$\alpha$} & 85 & 87 &  88 & 93 \\ 
\multicolumn{1} {|c|}{$\beta$} & ~5 & ~6 &  ~2 & ~4 \\ 
\multicolumn{1} {|c|}{$\phi$} & ~8 & ~9 & ~5 & ~8 \\ 
\multicolumn{1} {|c|}{$\epsilon$} & ~0 & ~1 & ~0 & ~1 \\  \hline
\end{tabular}
\caption{\footnotesize{Variance-based sensitivity indices for the $\sigma^2$ outputs}}\label{tab_sig}
\end{center}
\end{table}
\vspace{-0.5cm}

\begin{figure}[ht]
\begin{center}
\begin{tabular}{cc}
\includegraphics[scale=0.6,angle=0]{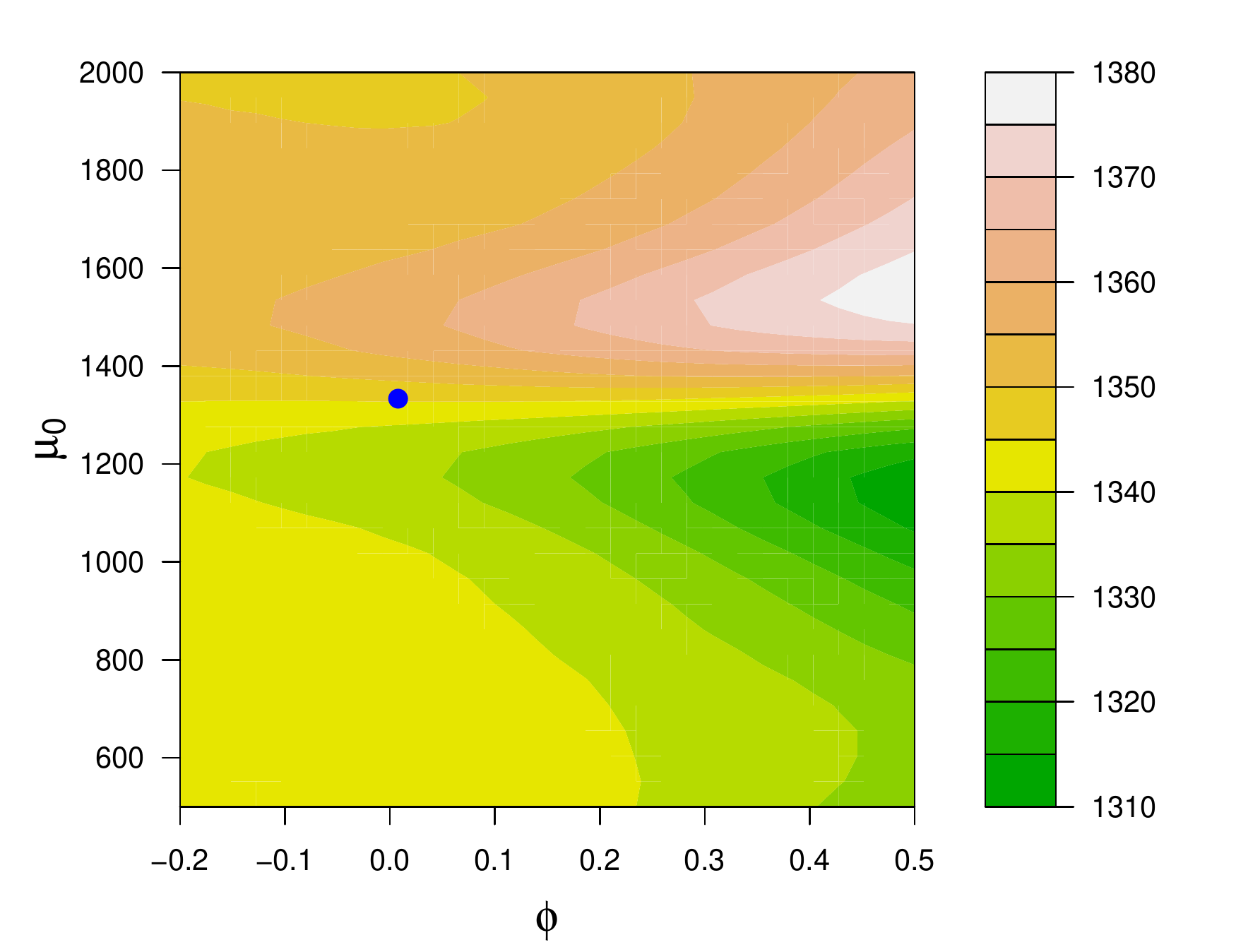}
\end{tabular}
\caption{\footnotesize{Joint effects plot showing the average effect of pairs of the inputs $\phi$ and $\mu_0$ on the analysis output of $\e{\mu|z}$.}}
\label{fig_full_inter}
\end{center}
\end{figure}

In addition to the plots in Figure~\ref{fig_full_me}, we are able to visualise the joint effect of two inputs by plotting the average value of the outputs conditioning on fixed values of two of the inputs.
The joint effect of $\mu_0$ and $\phi$ on  $\e{\mu|z}$ is shown in the  plot of Figure~\ref{fig_full_inter}: it is clear from that plot that the level of autocorrelation $\phi$, changes the influence of $\mu_0$, with larger positive values of $\phi$ resulting in a much stronger dependence on $\mu_0$. Again, these types of plots can be used to identify regions of the input space where the analysis is robust. Like for the plots of Figure~\ref{fig_full_me}, we could have presented these plots for any input-pair and output combination, but, for the most part, these plots were either flat, or just showed interesting behaviour in one dimension (which is represented by figure~\ref{fig_full_me}). 

On all the plots in Figures~\ref{fig_full_me} and \ref{fig_full_inter}, the location of the result of the original analysis from \cite{vicens1} is shown as a blue dot. By considering the plots and the variance-based sensitivity analysis results we can judge which outputs are robust to changes in which inputs within the vicinity of the original analysis. For instance, it is clear from these results that careful consideration needs to be given to the specification of $\mu_0$, $\phi$ and $\alpha$, and we know that the output E$(\mu|z)$ is particularly sensitive to changes in $\mu_0$ around the value used in the original analysis.

The emulator can also be used in a predictive manner: another scientist may come along who agrees with the original specification of $\mu_0$, $n_0$, $\alpha$ and $\beta$, but they believe that there is autocorrelation that is captured by setting $\phi=0.25$ and that the prior should be Cauchy rather than normal ($\epsilon=1$). The emulator can be queried to find immediately that, under this specification, we have the results in Table~\ref{tab_pred}, where we have an estimate of the relevant Bayesian analysis and an appreciation of the uncertainty caused by approximating the analysis using the emulator. We can see in this particular case that the posterior mean for $\mu$ is in the vicinity of the value from the original analysis (1,347), but the posterior variance for $\mu$ is far greater than the original (1,015), which is to be expected due to the increased correlation in the likelihood giving a reduced effective sample size, combined with a more uncertain prior. If this scientist was uncertain about the level of autocorrelation and wished to specify a distribution for $\phi$, the emulator could still be used to find their posterior summaries using the usual uncertainty analysis approach of \cite{oakoh2002}.

\begin{table}[ht]
\begin{center}
\begin{tabular}{|c|c|c|c|}\hline
Posterior  & Results from &Median & 90\% credible\\
summary &  \cite{vicens1}  & &  interval \\\hline
 $\e{\mu|z}$ & ~~1,347 &  			~~1,344&  (1,339,1,350)\\
$\var{\mu|z}$ & ~~1,015&			~~1,749& (1,424,2,073)\\
$\e{\sigma^2|z}$ &	61,937 &	68,210& (67,420,68,990)\\
$\var{\sigma^2|z}$  & 1.11$\times10^8$ &	1.47$\times10^8$ & (1.32$\times10^8$,1.62$\times10^8$)\\\hline
\end{tabular}
\caption{\footnotesize{Predictions from the emulator when using original values for $\mu_0$, $n_0$, $\alpha$ and $\beta$ alongside $\phi=0.25$ and $\epsilon=1$ (all results are to four s.f.).}}\label{tab_pred}
\end{center}
\end{table}

Of course, considering the sensitivity of the outputs in this way and performing predictions based on the emulator are just two of the ways we can investigate the original analysis using our method: we can also perform the type of analyses that were covered in Section~\ref{sec_ex_spec} and many more as discussed later in the present article. In particular, if there was interest in further exploring the robustness of the analysis in a particular part of input space, we could use the emulator to help select further MCMC runs for different inputs  in order to increase our knowledge of how the posterior summaries are affected in that region. The emulator-building exercise and the subsequent plotting of main and joint effects can also be useful as a diagnostic tool in that we may have prior beliefs about the way in which inputs influence the posterior outputs and these plots can help identify unexpected behaviour that may be due to programming bugs in the MCMC implementation.

\section{Discussion and future research directions}\label{sec_disc_cm}

\subsection{Modelling choices}\label{sec_disc_mod}

The proposed methodology of applying computer model emulation techniques to solve the global robustness or sensitivity 
problem for expensive Bayesian analyses raises many questions, some of which are related to those seen in a standard robust Bayesian analysis. We give our thoughts on these issues as follows (see \cite{insua1} for further discussion):

\bi
\item {\bf What should we vary?}  When setting up such analyses, we have many choices over what parts of the prior and likelihood we could vary. 
These can in the first instance be guided by the differing views held within the scientific community, or by our desire to test the sensitivity of our analysis to important parts of the specification. Key elicited quantities (such as the prior variance $\nu$ in Section~\ref{ssec_toy_mod_setup}) are obvious choices for parameters to vary, as are parameters representing relatively arbitrary but convenient assumptions, such as the 
contamination parameter $\epsilon$ from both sections~\ref{ssec_toy_mod_setup} and \ref{sec_riverflowiv}, which breaks conjugacy for $\epsilon>0$. In full generality, we may wish to vary everything possible, while maintaining consistency with the limited prior and likelihood specification. However, we should be careful here as although not explicitly stated, the prior specification may contain further reasonable but implicit structural information, such as 
unimodality and continuity of both the prior and likelihood pdfs, as well as additional, and possibly quite strict, bounds on the derivatives of the pdf's to ensure smoothness (the expert's beliefs, were we to interrogate them further, are unlikely to be jagged). This is critical as many robust analyses that leave out such additional constraints, can produce relatively non-informative results, especially in high dimension \citep[this links to arguments made by][]{Gustafson:1995aa}.
These constraints therefore greatly restrict the class of analyses we should use, and hence may allow parameterised approaches, such as we present here, to capture the major sources of variation.  
The limitations as to what we can vary link to the concept of {\it model discrepancy} that we discuss in Section~\ref{ssec_future}, which would 
capture the additional uncertainty that our current representation ignores.

\item {\bf How do we decide how to contaminate a prior or likelihood?}
As we have demonstrated, the contamination of a prior or likelihood represents a simple to implement parameterised method of breaking away from mathematically convenient distributional assumptions, while respecting core scientific principles. There is of course much freedom in the choice of contaminating distribution, however, we would usually want to ensure the contaminant possesses key attributes found in the  uncontaminated term. For example, in Section~\ref{ssec_toy_mod_setup} the contaminant to the likelihood 
$\pi(z_i|\theta,\epsilon=1)$ given in equation~(\ref{eq_contam}) was chosen to have the same expectation of $1/\theta$, as the uncontaminated term $\pi(z_i|\theta,\epsilon=0)$ given by equation~(\ref{eq_toy_conj2}). Similarly the prior contamination of Section~\ref{ssec_ext_con_ana}, shared the same quartiles as the original prior specification. Note that in both cases the contaminants 
shared with the original analysis the additional properties of unimodality and continuity of the pdfs (and derivatives). While these perturbations 
are by their very nature limited, it would still be comforting to find that the key features of the posterior, or end decision process, are robust to them,
and highly informative to find the opposite.

\item {\bf What should the exploratory space look like? In the simplest case, what ranges should we use?}
Ideally, the exploratory space $\mathcal{X}$ should contain the differing specifications that exist across all, or at least some specific subset, of the relevant scientific community. Note that $\mathcal{X}$ may contain regions $\mathcal{X}_k$ representing individual robust Bayesian analyses that scientists wish to perform, such as cases 2, 3 and 4 in section~\ref{sec_ex_spec}. While in practice this would be difficult to achieve precisely, as the scientific community may not agree exactly with our choice of parameterisation, we would still hope to capture the major aspects of the differences of opinion across the area. 
This implies that there is an important difference between $\mathcal{X}$ and the corresponding region $\mathcal{X}_k$ explored in a single perfunctory robust Bayesian analysis, in that $\mathcal{X}$ just needs to cover all areas of interest, and assuming it achieves this, the precise location of its boundary is of somewhat less importance (however, although in principal our proposed emulation methodology can deal with large numbers of inputs defined over wide ranges, the smaller $\mathcal{X}$ is, the easier it may be to emulate).
In contrast, when specifying a particular region 
$\mathcal{X}_k$ for use in a robust Bayesian analysis, where interest lies in the extrema of $f(x)$ over $\mathcal{X}_k$, such as in cases 2 to 4 in Section~\ref{sec_ex_spec}, the geometry and extent of the boundaries of $\mathcal{X}_k$ should be considered very carefully. For example, often, $\mathcal{X}_k$ may be constructed from the intersection of univariate interval constraints on the components of $x$, implying $\mathcal{X}_k$ is a hypercube. However, this is usually just a convenient construction, and can posses disadvantages: 
as $f(x)$ may display noticeably different behaviour in the many corners of such a hypercube the corners may dominate the robust analysis. An elliptical specification, as used in case 4 in Section~\ref{sec_ex_spec}, may be both more realistic and simultaneously easier to emulate. 
These issues have caused problems in previous robustness studies in differing dimensions: while exploring wide classes of priors in 1-dimension can still lead to meaningful conclusions~\citep{berger1}, in higher dimensions such artificial classes of priors can overwhelm the data, leading to non-informative results \citep{insua1}, a problem that will become worse when we simultaneously perturb the likelihood too.



\item {\bf What happens if we cannot emulate the Bayesian analysis?} One can envisage a particularly erratically behaved Bayesian analysis where standard emulation procedures would perform poorly, as would most likely be flagged by emulator diagnostics~\citep{bastos1}. In this case, we would a) be very glad to have been made aware of this erratic behaviour across $\mathcal{X}$ and b) most likely suggest the analysis would fail any reasonable test of robustness. Hence if we cannot emulate it, we would be unlikely to trust it. We may then attempt to emulate sub-components of the full analysis, to investigate its structure further and to identify the cause of the non-robust behaviour.

\item {\bf What can we do if we cannot assume smoothness?} If we are uncomfortable with the standard smoothness assumption across $\mathcal{X}$, we can 
use alternative forms for the emulator correlation function that represent non-smooth surfaces with particular attributes. If we suspect the output to have sudden  discontinuities, either in its derivative or in the function $f(x)$ itself, we can attempt to identify the location of such discontinuities using history matching techniques discussed below. 

\ei

\subsection{Future Research Directions}\label{ssec_future}

The proposed methodology raises several possible future research directions. For example, there are powerful computer model techniques that have interesting analogies in this context:

\bi

\item {\bf History Matching}: say we were interested in identifying a subset $\mathcal{X}_0 \subset \mathcal{X}$ of the class of specifications that satisfied some criteria 
on the posterior, possibly related to a downstream decision calculation. We could do an initial wave of runs as presented here,
to emulate the posterior features of interest across the whole space, and rule out regions of $\mathcal{X}$ where we are sure the criteria would 
not be satisfied, taking 
into account uncertainties as represented by the emulator variance. We could then perform successive waves of runs focussed on the regions where possible matches of the criteria occur, designed to reduce the emulator variance and hence to identify if such a region exists and its precise location in $\mathcal{X}$.
This is the analogy in this context, to history matching: a computer model technique that has proved very successful 
across a wide range of scientific disciplines, including cosmology~\citep{Vernon10_CS,Vernon10_CS_rej,vernon_astro,
galf_stat_sci,Vernon:2016aa}, epidemiology~\citep{Yiannis_HIV_1,Yiannis_HIV_2}, oil reservoir modelling~\citep{Craig96_Pressure,Craig97_Pressure,JAC_Handbook,JAC_sma_samp}, climate modelling~\citep{Williamson:2013aa}, 
environmental science~\citep{asses_mod} and systems biology~\citep{Vernon_sysbio_hm_2016}.

\item {\bf Model discrepancy}: a key feature of current computer model analyses is the inclusion of a model discrepancy term~\citep[see e.g.][]{Craig97_Pressure,kenoh2001,Brynjarsdottir:2014aa}. This is an 
upfront acknowledgement of the deficiencies of a scientific computer model due to missing physics, simplifying assumptions, imperfect solvers etc. 
In our current context of a Bayesian analysis, the model discrepancy would represent the simplifying assumptions used throughout the construction of the 
Bayesian model and prior specification, including use of convenient mathematical forms for distributions, the fixing of various parameters, simplifications made to the model's structure (say for mathematical convenience or due to time constraints on the analysis), and 
the limited belief specifications of the expert.
It is currently somewhat hypocritical to criticise a scientist for excluding such a term in their analysis, 
when in our Bayesian statistics community the vast majority of analyses do little better. 
Of course, a major motivation for performing a sensitivity/robust analysis is precisely to try to identify and characterise some of the 
dominant deficiencies in the Bayesian 
calculation, but this process is {\it necessarily incomplete}, as we cannot hope to parameterise all the possible alterations to the model we would wish to explore. The model discrepancy should capture the remaining uncertainty that we have left out of the parameterisation and would therefore link our current robust analysis 
with the robust analysis that we would wish to do given more time, computational resources and expert input. It would essentially summarise how much we should trust this robust Bayesian analysis. There is much more investigation of this concept required, but see \cite{Goldstein_10}
for an argument basing the need for model discrepancy as applied to an expectation based Bayesian analysis on foundational principles such as temporal sure preference, and see also \cite{D:2015aa} for a description of the novel but related technique of posterior belief assessment.

\ei
Other lines of research more specific to this context are also possible:

\bi

\item {\bf Structured Emulator Priors}: 
There are several situations where we would have detailed insight into the result of the Bayesian update for specific subsets of $\mathcal{X}$. Any known 
structural knowledge could then be incorporated into an informed prior for the emulator, leading to a possibly substantial reduction in the number of 
MCMC evaluations we need to cover the input space. The simplest example is when $\mathcal{X}$ includes a surface where the prior variance of a parameter $\theta$ is zero. We hence know the posterior features of $\theta$ are the same as the prior, and this can be built into the emulator. A more useful case is when $\mathcal{X}$ contains a surface where a conjugate analysis is possible, such as is true for both the examples in this paper (e.g. the $x$-axis in figure~\ref{fig_toy_emul}, both panels). Here there will be highly structured 
information that can be included in the emulator prior for minimal computational cost using ``known boundary emulation" techniques \citep{Vernon:aa}. In addition, structured priors can be formed from fast but approximate methods of evaluating the posterior, in direct analogy to multilevel emulation \citep[see e.g. ][]{Craig97_Pressure,JAC_sma_samp}.
In fact, depending on the posterior features of interest, there is a rich hierarchy of informed priors one could use, but we leave this to future work \citep{Vernon:2017aa}. 
It is worth noting, however, that in this scenario we elicit priors for the emulator from the statistician, not the subject matter expert, as the statistician is the expert when it comes to understanding the structure of a Bayesian update. Hence, in general, detailed elicitations may be possible.

\item {\bf MCMC development}: We envisage improvements to MCMC algorithms tailored to this type of analysis. A major criticism of 
our approach is that current MCMC calculations are often extremely expensive taking days, weeks or even months, and 
the requirement to perform $n$ evaluations over $\mathcal{X}$ is therefore prohibitively expensive. However, MCMC is notoriously hard to parallelise; in contrast to our approach where parallelisation is trivial: a user with a cluster of $n$ or greater processors will see no
increase in wall clock time by performing this form of analysis. If the runs must be done in sequential batches however, these should be ordered efficiently by exploiting the geometry of the design, 
in that the end point of one MCMC chain located at $x_1\in \mathcal{X}$ 
would make an ideal initial condition for a chain located at $x_2\in \mathcal{X}$ were $|x_1 - x_2|$ considered small, hence reducing burn in time. 
Similar efficiency savings can be made with parameters governing the adaptation of the MCMC algorithm, as we can again reuse those from completed neighbouring runs. 
MCMC convergence
criteria could also be improved for batches of similar runs. Most importantly, methods for exploring a limited 
sensitivity analyse to prior parameters by evaluating a diffuse version of the 
posterior using a single MCMC run, could also be extended and used to update the emulator using runs that now represent 
low dimensional subspaces of $\mathcal{X}$ (e.g. lines or hyperplanes, spanning all or part of $x_p$), instead of single points. One may be tempted to extend this approach further, and
to design a single MCMC run that explores all of $\mathcal{X}$ as well as $\Theta$ (the support of $\theta$) simultaneously. However this may be an inefficient approach as the combined space $\mathcal{X} \times \Theta$ would be of much higher dimension than just $\Theta$, most likely possessing a far more complex structure, and hence greatly extending MCMC convergence times, compared to the approach we use here. 
Perhaps of most use would be to incorporate MCMC based local sensitivity analysis \citep[][]{Perez:2005aa} into the emulator via the derivative structure given by equation~(\ref{eq_mcderivcov}), an approach that we again leave to future work \citep{Vernon:2017aa}.
Finally, if we have an a priori estimate of the length of burn in and the amount of thinning that may be required (that is, the effective sample size per MCMC step) we can perform more advanced 
design calculations to determine the optimal number $n$ of locations in input space to run the MCMC algorithm for a fixed computational budget, to optimise emulator performance. 



\ei

\section{Conclusion}\label{sec_conc}

In this article we have proposed a framework for addressing the general Bayesian robustness problem, in which we treat complex and computationally demanding Bayesian analyses as expensive computer models. We applied emulation technology developed for complex computer models to explore the structure of the Bayesian analysis itself, and, specifically, its response to various changes in both the prior and likelihood specification. This allows for a more general sensitivity and robustness analysis, and provides a very flexible methodology that could in addition be applied to a wide class of statistical analyses. 

It could be argued that every important Bayesian analysis, where the results may have serious consequences, should employ a robust 
analysis of the kind we propose here. Enabling the analysis of classes of prior and likelihood specification should also help the uptake of Bayesian methods within scientific communities, as each expert will have access to the posterior attributes corresponding to their own subjective beliefs. Experts may also find 
the answer to the question: ``how far do I have to perturb my specification before my decision changes" to be easy to interpret, and help assuage their fears 
over the use of specific choices of subjective priors and likelihood, due to their (possible) robustness. In many contexts this strategy may be of more use that the
standard Bayesian approach of adding another layer to the model hierarchy, requiring the assertion of possibly artificial priors over the space $\mathcal{X}$, the meaning of which may be questionable. 

There are now some extremely expensive MCMC algorithms that may take weeks or longer to run, and hence there may be obvious practical challenges that make it impossible to perform repeated evaluations as required for this analysis. However we should insist on turning this around, and our response would be the same as is often given to climate scientists, who also construct extremely expensive models: if the model is too expensive to
allow a reasonable sensitivity analysis, why should we trust in its results at all? Such considerations would promote a welcome change in emphasis in Bayesian statistics away from extremely complex models and algorithms, and toward well understood, robust and trustworthy analyses. This, combined with further developments to tailor MCMC algorithms for efficient use in this context, maybe a sensible direction for Bayesian statistics to take.






\section*{Acknowledgements}

We thank Peter Craig and Richard Wilkinson for valuable discussions when we were researching our method and related topics.





\bibliographystyle{ba}
\bibliography{BA2}


\vspace{-0.5cm}
\section*{Appendix A}
\vspace{-0.2cm}
\begin{table}[h]
\begin{center}
\begin{tabular}{|l|llllllllll|}
\hline 
$i$ &1&2&3&4&5&6&7&8&9&10 \\
\hline
$z_i$ &1.169 &0.386 &1.164 &0.028 &0.506 &0.287 &0.911 &0.200 &0.289& 0.381 \\
\hline 
\end{tabular}
\caption{\footnotesize{The simulated data used in the example Bayesian analysis of section~\ref{ssec_toy_mod_setup}.}}
\end{center}
\end{table}

\end{document}

%% file: NewPlots/Table1_altered.tex
\begin{table}[t]
\centering
\begin{tabular}{|r|rrr|rrr|}
  \hline
 &  $f_1(x)$ & $\frac{\partial f_1(x)}{\partial \nu}$ & $\frac{\partial f_1(x)}{\partial \epsilon}$ & $f_2(x)$ & $\frac{\partial f_2(x)}{\partial \nu}$ & $\frac{\partial f_2(x)}{\partial \epsilon}$ \\ 
  \hline
$\e{\, .\,}$ & 2.596 & -0.693 & -0.423 & 0.533 & -0.048 & -0.232 \\ 
  $\SD{\, .\,}$ & 0.020 & 0.111 & 0.178 & 0.003 & 0.015 & 0.024 \\ 
   \hline
\end{tabular}
\caption{\footnotesize{Results of the local sensitivity analyis corresponding to specification case 1. The first row gives the emulator expectation of all requested quantities of interest, namely the posterior mean $f_1(x)$, posterior SD $f_2(x)$ and partial derivatives of each. The second row gives the corresponding emulator SD of each of these estimates, which could be reduced using further MCMC runs.}} 
\label{tab_case1}
\end{table}